\documentclass[a4paper,11pt]{article}

\usepackage{jheppub} 

\usepackage[T1]{fontenc} 

\usepackage{latexsym}
\usepackage{amsmath,amsfonts}
\usepackage{times}
\allowdisplaybreaks[4]

\def\nline{\,\nabla\kern -0.7em\raise0.2ex\hbox{/}\,\,}
\def\yline{\,y\kern -0.47em /}
\def\aline{\,a\kern -0.49em /}
\def\parline{\,\partial\kern -0.55em /\,\,}

\newcommand{\Co}{\mathbb{C}}

\newcommand{\No}{\mathbb{N}}

\newcommand{\Ro}{\mathbb{R}}

\def\be{\begin{equation}}
\def\ee{\end{equation}}
\def\beq{\begin{eqnarray}}
\def\eeq{\end{eqnarray}}
\def\beqq{\begin{eqnarray*}}
\def\eeqq{\end{eqnarray*}}

\def\CSFsm{{\scriptscriptstyle \rm CSF}}
\def\ISFsm{{\scriptscriptstyle \rm ISF}}
\def\csfsm{{\scriptscriptstyle \rm csf}}
\def\isfsm{{\scriptscriptstyle \rm isf}}

\def\smone{{\scriptscriptstyle (1)}}

\def\smp3{{\scriptscriptstyle [3]}}

\def\Cbf{{\bf C}}
\def\Jbf{{\bf J}}

\def\Qbf{{\bf Q}}

\def\ibf{{\bf i}}
\def\iibf{{\bf ii}}
\def\iiibf{{\bf iii}}
\def\ivbf{{\bf iv}}

\def\LL{{\cal L}}
\def\NN{{\cal N}}

\def\SSc{{\cal S}}

\def\vph{{\vphantom{5pt}}}

\def\half{\frac{1}{2}}

\def\Gb{{\bar{G}}}

\def\alphab{\bar{\alpha}}

\def\zetab{\bar{\zeta}}

\def\ccc{c}

\def\Thsm{{\scriptscriptstyle \rm Th}}

\def\irm{{\rm i}}

\def\msv{{\rm msv}}
\def\msl{{\rm msl}}

\def\Erm{{\rm E}}
\def\EPLrm{{\rm EPL}}
\def\PLrm{{\rm PL}}
\def\covrm{{\rm  cov}}
\def\locrm{{\rm  loc}}
\def\lcrm{{\rm  lc}}
\def\lcfrm{{\rm  lcf}}
\def\lcgrm{{\rm  lcg}}
\def\offshrm{{\rm  off-sh}}

\def\noinbf#1{\noindent {\bf #1}}

\title{\boldmath  Lorentz covariant on-shell cubic vertices for
continuous-spin fields and integer-spin fields}

\author[a]{R.R. Metsaev}

\affiliation[a]{Department of Theoretical Physics, P.N. Lebedev Physical
Institute, \\ Leninsky prospect 53,  Moscow 119991, Russia}

\emailAdd{metsaev@lpi.ru}

\abstract{In the framework of Lorentz covariant on-shell approach, interacting continuous-spin fields and integer-spin fields in flat space are investigated. Continuous-spin fields are considered by using a Lorentz vector superspace formulation, while integer-spin fields are considered by using an oscillator formulation.
All parity-even cubic vertices for self-interacting continuous-spin fields realized as functions on the Lorentz vector superspace are obtained. Cross-interactions of continuous-spin fields and integer-spin fields are also derived. Several representatives of cubic vertices realized as distributions are  obtained.
We show that manifestly Lorentz invariant formal cubic action involving at least one continuous-spin field turns out to be divergent. We find the modification of such action which maintains Lorentz invariance and leads to finite cubic action. One-to-one correspondence of Lorentz covariant cubic vertices and light-cone gauge cubic vertices is demonstrated explicitly.

\vspace{1cm}

\keywords{Lorentz covariant vector superspace, continuous and higher spin fields, interaction vertices}.

\vspace{1cm}

\arxivnumber{2510.05011 V2 [hep-th]\,, \qquad Preprint FIAN-TD-19}

}

\begin{document}
\maketitle
\flushbottom

\section{Introduction}

Continuous-spin field theory is one of the interesting topics in modern quantum field theory. For review of a continuous-spin field (CSF) and extensive list of references, see refs.\cite{Bekaert:2006py,Brink:2002zx,Bekaert:2017khg}. The pioneering contributions in the development of a Lagrangian formulation for free bosonic and fermionic CSFs may be found in the respective refs.\cite{Schuster:2013pta} and ref.\cite{Najafizadeh:2015uxa}. Interacting CSFs have been attacked by using a field theoretical approach and a worldline approach in  refs.\cite{Metsaev:2017cuz,Bekaert:2017xin,Rivelles:2018tpt, Metsaev:2018moa,Rivelles:2023hzo,Metsaev:2025qkr,Schuster:2023xqa}. Namely, in the framework of a light-cone gauge field theoretical approach, interacting CSFs were considered in refs.\cite{Metsaev:2017cuz,Metsaev:2018moa,Metsaev:2025qkr}, while a Lorentz covariant field theoretical approach to CSFs was investigated in refs.\cite{Bekaert:2017xin,Rivelles:2018tpt,Rivelles:2023hzo}. The extensive study of interacting CSFs by using a worldline approach may be found in refs.\cite{Schuster:2023xqa}.
Attempts for S-matrix formulation of CSF have been reported in refs.\cite{Schuster:2013pxj,Bellazzini:2024dco}, while thermodynamical aspects of CSF have been investigated in ref.\cite{Schuster:2024wjc}.

In refs.\cite{Metsaev:2017cuz,Metsaev:2018moa}, we developed the oscillator light-cone gauge formulation of  massive/massless CSFs and applied such formulation for study of cubic vertices for CSFs. Unfortunately  the oscillator light-cone gauge formulation leads to somewhat complicated cubic vertices expressed in terms of special functions (Bessel and hypergeometric functions) and deals with complicated dressing operators acting on the special functions. This motivated us to develop the alternative light-cone gauge formulations in ref.\cite{Metsaev:2025qkr} which we refer to as the light-cone gauge vector superspace formulation. As shown in ref.\cite{Metsaev:2025qkr} the vector formulation, in contrast to the oscillator formulation, leads to simple cubic vertices expressed in terms of rational or exponential functions.

Our aim in this paper is to find a Lorentz covariant formulation of the cubic vertices for CSFs obtained in the framework of light-cone gauge vector formulation in ref.\cite{Metsaev:2025qkr}. There are the following three reasons for our interest in the Lorentz covariantization of the cubic vertices in ref.\cite{Metsaev:2025qkr}.

\noinbf{i)} The Lorentz covariant on-shell formulation of CSFs is based in the use of relativistic fields subject to constraints proposed by Wigner for massless CSF (W-constraints) and Boulanger and Bekaert for massive CSF (BB-constraints). For review of the constraints, see ref.\cite{Bekaert:2006py}. Though there are various point of views and interesting alternative formulations, we think that relativistic fields  subject to WBB-constraints provide the reliable, trustable, and basic description of CSFs. Therefore it is desirable to get Lorentz covariant description for cubic vertices directly in terms of the relativistic fields subject to WBB-constraints.

\noinbf{ii)} In order to study Lorentz covariant cubic vertices we use so called Lorentz covariant on-shell approach. Such approach, on the one hand, is closely related to the light-cone gauge approach and, on the other hand, provides good starting point for a study of various Lorentz covariant off-shell formulations.

\noinbf{iii)} As compared to the light-cone approach, the Lorentz-covariant approach is more transparent and easy to reproduce and hence could be useful for a more wide audience.

In order to describe a general classification of cubic vertices let us use the following shortcuts for continuous-spin field (CSF) and integer-spin field (ISF) propagating in the flat space $\Ro^{d-1,1}$:
{\small
\beqq
\begin{array}{ll}
(m,\SSc)_\CSFsm, \ \  m^2 < 0, \ \ \SSc\in \Co,   \ \ \hbox{for massive CSF}; \hspace{0.5cm} & (m,s)_\ISFsm, \ \ m^2 > 0,  \ s\in \No_0, \ \hbox{for massive ISF};
\\[7pt]
(0,\kappa)_\CSFsm,  \ \ \ \kappa^2 > 0 \hspace{2cm} \hbox{for massless CSF};  &  (0,s)_\ISFsm, \ \ s\in \No_0, \hspace{1.7cm} \hbox{for massless ISF}.\qquad
\end{array}
\eeqq
}
In this paper, we consider the massive CSF with the value of generalized spin $\SSc$ given by $\SSc= \frac{3-d}{2}+q$, $q^*=-q$. Such $\SSc$ corresponds to so-called principal series. For ISF, the label $s\in \No_0$ stands for value of spin.  Solutions for cubic vertices realized
as functions on the Lorentz vector superspace are referred to as {\it f-solutions}, while solution for cubic vertices realized as distributions are referred to as {\it d-solutions}.
The general classification of a priori non-trivial cubic vertices involving at least one massive/massless CSF, which we borrow from ref.\cite{Metsaev:2025qkr}, can be presented in the following way:
{\small
\beq
&& \hspace{-3cm} \hbox{\bf Cubic vertices:}
\nonumber\\
&& \hspace{-2.5cm} \hbox{\bf Three continuous-spin fields:}
\nonumber\\
\label{15092025-man03-01} && \hspace{-2.5cm} (m_1,\SSc_1)_\CSFsm\hbox{-}(m_2,\SSc_2)_\CSFsm\hbox{-}(m_3,\SSc_3)_\CSFsm\,, \hspace{4cm} \hbox{f-solution}
\\
\label{15092025-man03-02} && \hspace{-2.5cm} (m_1,\SSc_1)_\CSFsm\hbox{-}(m_2,\SSc_2)_\CSFsm\hbox{-}(0,\kappa_3)_\CSFsm\,, \hspace{0.5cm} m_1^2 \ne m_2^2 \,;    \hspace{2cm} \hbox{f-solution}
\\
\label{15092025-man03-03} && \hspace{-2.5cm} (m_1,\SSc_1)_\CSFsm\hbox{-}(m_2,\SSc_2)_\CSFsm\hbox{-}(0,\kappa_3)_\CSFsm\,, \hspace{0.5cm}  m_1^2=m_2^2\,; \hspace{2cm} \hbox{f-solution}
\\
\label{15092025-man03-04} && \hspace{-2.5cm} (m_1,\SSc_1)_\CSFsm\hbox{-}(0,\kappa_2)_\CSFsm\hbox{-}(0,\kappa_3)_\CSFsm\,,  \hspace{4.7cm} \hbox{f-solution}
\\
\label{15092025-man03-05} && \hspace{-2.5cm} (0,\kappa_1)_\CSFsm\hbox{-}(0,\kappa_2)_\CSFsm\hbox{-}(0,\kappa_3)_\CSFsm\,,  \hspace{5cm}  \hbox{d-solution, no f-solution}
\\[10pt]
&& \hspace{-2.5cm}  \hbox{\bf Two continuous-spin fields and one integer-spin field:}
\nonumber\\
\label{15092025-man03-06} && \hspace{-2.5cm} (m_1,\SSc_1)_\CSFsm\hbox{-}(m_2,\SSc_2)_\CSFsm\hbox{-}(m_3,s_3)_\ISFsm\,,  \hspace{4cm} \hbox{f-solution}
\\
\label{15092025-man03-07}  && \hspace{-2.5cm} (m_1,\SSc_1)_\CSFsm\hbox{-}(m_2,\SSc_2)_\CSFsm\hbox{-}(0,s_3)_\ISFsm\,, \hspace{0.5cm}  m_1^2\ne m_2^2\,;  \hspace{2cm} \hbox{f-solution}
\\
\label{15092025-man03-08}  && \hspace{-2.5cm} (m_1,\SSc_1)_\CSFsm\hbox{-}(m_2,\SSc_2)_\CSFsm\hbox{-}(0,s_3)_\ISFsm\,,  \hspace{0.5cm}    m_1^2 =  m_2^2\,;  \hspace{2cm} \hbox{f-solution}
\\
\label{15092025-man03-09}  && \hspace{-2.5cm} (m_1,\SSc_1)_\CSFsm \hbox{-}(0,\kappa_2)_\CSFsm\hbox{-}(m_3,s_3)_\ISFsm\,,   \hspace{4.2cm} \hbox{f-solution}
\\
\label{15092025-man03-10} && \hspace{-2.5cm} (m_1,\SSc_1)_\CSFsm\hbox{-}(0,\kappa_2)_\CSFsm\hbox{-}(0,s_3)_\ISFsm\,,  \hspace{4.5cm} \hbox{f-solution}
\\
\label{15092025-man03-11} && \hspace{-2.5cm} (0,\kappa_1)_\CSFsm\hbox{-}(0,\kappa_2)_\CSFsm\hbox{-}(m_3,s_3)_\ISFsm\,,  \hspace{4.5cm} \hbox{f-solution}
\\
\label{15092025-man03-12} &&  \hspace{-2.5cm} (0,\kappa_1)_\CSFsm\hbox{-}(0,\kappa_2)_\CSFsm\hbox{-}(0,s_3)_\ISFsm\,,  \hspace{4cm}       \hspace{0.8cm} \hbox{d-solution, no f-solution}
\\[10pt]
&& \hspace{-2.5cm}  \hbox{\bf One continuous-spin field and two integer-spin fields:}
\nonumber\\
\label{15092025-man03-13}  && \hspace{-2.5cm} (m_1,\SSc_1)_\CSFsm^\vph\hbox{-}(m_2,s_2)_\ISFsm\hbox{-}(m_3,\kappa_3)_\ISFsm\,,  \hspace{4cm} \hbox{f-solution}
\\
\label{15092025-man03-14} && \hspace{-2.5cm} (m_1,\SSc_1)_\CSFsm^\vph\hbox{-}(m_2,s_2)_\ISFsm\hbox{-}(0,s_3)_\ISFsm\,,  \hspace{4.4cm} \hbox{f-solution}
\\
\label{15092025-man03-15} &&  \hspace{-2.5cm} (m_1,\SSc_1)_\CSFsm^\vph\hbox{-}(0,s_2)_\ISFsm\hbox{-}(0,s_3)_\ISFsm\,,  \hspace{4.7cm} \hbox{f-solution}
\\
\label{15092025-man03-16} && \hspace{-2.5cm} (0,\kappa_1)_\CSFsm \hbox{-}(m_2,s_2)_\ISFsm\hbox{-}(m_3,s_3)_\ISFsm\,,  \hspace{0.6cm} m_2^2  \ne m_3^2\,;  \hspace{2cm} \hbox{f-solution}
\\
\label{15092025-man03-17} && \hspace{-2.5cm} (0,\kappa_1)_\CSFsm\hbox{-}(m_2,s_2)_\ISFsm\hbox{-}(m_3,s_3)_\ISFsm\,, \hspace{0.6cm} m_2^2 = m_3^2\,,\hspace{1cm}  \hspace{1cm} \hbox{d-solution?, no f-solution}
\\
\label{15092025-man03-18} && \hspace{-2.5cm} (0,\kappa_1)_\CSFsm\hbox{-}(m_2,s_2)_\ISFsm\hbox{-}(0,s_3)_\ISFsm\,,   \hspace{4.6cm} \hbox{f-solution}
\\
\label{15092025-man03-19} && \hspace{-2.5cm} (0,\kappa_1)_\CSFsm\hbox{-}(0,s_2)_\ISFsm\hbox{-}(0,s_3)_\ISFsm\,,   \hspace{4.9cm} \hbox{no f-solution}
\eeq
}
For the reader convenience we now briefly summarize our results obtained in this paper.

\noinbf{i}) In \eqref{15092025-man03-01}-\eqref{15092025-man03-19} we show explicitly all cubic vertices which admit (or do not admit) f-solutions.

\noinbf{ii}) For cubic vertices \eqref{15092025-man03-05}, \eqref{15092025-man03-12}, we find some particular d-solutions. Note however that, at the present time, we are not aware a self-contained and systematic method for finding all d-solutions. This is to say that finding all d-solutions is beyond the scope of our paper.

\noinbf{iii}) We demonstrate that our f- and d-solutions for the Lorentz covariant cubic vertices are in one-to-one correspondence with the ones for light-cone gauge cubic vertices in ref.\cite{Metsaev:2025qkr}.

\noinbf{iv)} For cubic vertex in \eqref{15092025-man03-17}, we found the light-cone gauge d-solution in ref.\cite{Metsaev:2025qkr}. At the present time, we are not aware of a Lorentz covariant counterpart of such light-cone gauge d-solution. Note however that we have no proof that a Lorentz covariantization of our light-cone gauge d-solution  for cubic vertex \eqref{15092025-man03-17} obtained in ref.\cite{Metsaev:2025qkr} is not possible.

\noinbf{v)} We find all f-solutions for parity-even cubic vertices for the case of $d>4$, where $d$ stands for the dimension of the flat space $\Ro^{d-1,1}$. To avoid unnecessary technical complications and too long discussion the case $d=4$ will be separately considered elsewhere. In this paper we prefer not to deal with particular properties of cubic vertices in $d=4$ and present vertices which work on equal footing for all dimensions $d\geq 4$.

The presentation in our paper is organized in the following way. Section \ref{sec-02} is devoted to review of the Lorentz covariant on-shell formulation of CSF and ISF.

In section \ref{sec-03}, we present our result for manifestly Lorentz invariant formal cubic action. Going to the light-cone frame we show explicitly divergencies of the formal action. We propose the modification of the formal action that maintains of the Lorentz invariance and remove the divergencies.

Section \ref{sec-3csf} is devoted to Lorentz covariant cubic vertices for CSFs. We present cubic vertices for self-interactions of massive/massless CSFs and cubic vertices for cross-interactions of massive CSFs and  massless CSFs.  Lorentz covariant cubic vertices for cross interactions of CSFs and ISFs are discussed in sections \ref{sec-2csf}, \ref{sec-1csf}. Also in sections \ref{sec-3csf}-\ref{sec-1csf}, we demonstrate explicitly the one-to-one correspondence of on-shell Lorentz covariant cubic vertices and light-cone gauge cubic vertices.

In section \ref{concl}, we present our conclusions. Notation and conventions used in this paper are presented in appendix \ref{app-notation}.
In appendix \ref{div}, we present basic relations of light-cone gauge formulation of CSF and ISF we use in this paper. In Appendices \ref{proof}, \ref{lor-inv}, we discuss some technical details of the proof of Lorentz invariance of the cubic action we propose. In appendix \ref{connect}, we describe a connection between Lorentz covariant cubic vertices and their light-cone gauge cousins.

\section{Free Lorentz covariant continuous-spin fields and integer-spin fields}\label{sec-02}

We start with the brief review of the Lorentz covariant on-shell formulation of CSF and ISF. For massive CSF we use formulation proposed by Bekaert and Boulanger, while, for massless CSF, we use formulation proposed by Wigner. For more discussions about these formulations, see ref.\cite{Bekaert:2006py}. For massive/massless ISFs we use old fashioned Lorentz covariant and gauge invariant on-shell formulation.

\noinbf{Massive/massless continuous-spin field}. To discuss massive/massless CSF we introduce the field $\phi= \phi(p,\xi)$ which depends on momentum $p^\mu$ and Lorentz $so(d-1,1)$ algebra vector $\xi^\mu$. By definition
the momentum $p^\mu$ and the vector $\xi^\mu$ are constrained to the surface defined as%
\footnote{Relation of the constraints in \eqref{16092025-01-man03-01} to the original form of WBB-constraints is clarified in appendices B, C in ref.\cite{Metsaev:2025qkr}. Various modifications of W-constraints may be found in refs.\cite{Najafizadeh:2017tin}.}
\be \label{16092025-01-man03-01}
\begin{array}{llll}
p^\mu p^\mu + m^2 = 0, \hspace{0.5cm} &   p^\mu \xi^\mu   = 0, \hspace{0.5cm} &  \xi^\mu \xi^\mu   = 0 \,, & \hbox{for massive CSF};
\\[9pt]
p^\mu p^\mu = 0,     & p^\mu \xi^\mu = 0, & \xi^\mu\xi^\mu  - \kappa^2 = 0, \hspace{0.5cm} & \hbox{for massless CSF};
\end{array}
\ee
where, for massive CSF, $m^2<0$, while, for massless CSF, $\kappa^2>0$. The CSF $\phi(p,\xi)$ defined on the surface \eqref{16092025-01-man03-01} obeys the constraints
\be \label{16092025-01-man03-02}
\begin{array}{ll}
\big( \xi^\mu \partial_{\xi^\mu} - \SSc \big) \phi = 0, \quad \SSc := \frac{3-d}{2}+q, \quad q^*=-q\,, \hspace{0.5cm} & \hbox{for massive CSF};
\\[9pt]
\big(p^\mu\partial_{\xi^\mu} -\irm\big)\phi  = 0, & \hbox{for massless CSF};
\end{array}
\ee
where $\partial_X:= \partial/\partial X$. As side remark, there are no gauge symmetries in the description of massive/massless CSFs we use throughout this paper.

\noinbf{Massive/massless integer-spin fields and triplet integer-spin fields}. In the framework of the on-shell Lorentz covariant and gauge invariant approach, massive/massless (triplet) ISFs are described by the following fields defined in a momentum space:
\be \label{16092025-01-man03-10}
\begin{array}{lll}
\bigoplus_{n=0}^s\,\,\phi^{\mu_1\ldots \mu_n}(p),  \hspace{0.5cm} & p^\mu p^\mu + m^2 =0, \hspace{0.5cm} &   \hbox{for massive (triplet) ISF};
\\[9pt]
\phi^{\mu_1\ldots \mu_s}(p),  & p^\mu p^\mu = 0 &  \hbox{for massless (triplet) ISF};
\end{array}
\ee
where, for massive (triplet) ISF, $m^2>0$. For massive (triplet) ISF in \eqref{16092025-01-man03-10}, the fields $\phi^{\mu_1\ldots \mu_n}(p)$ with $n=0,1$, and $n\geq 2$ stand for the respective scalar, vector, and rank-$n$ totally symmetric tensor fields of the Lorentz algebra $so(d-1,1)$. For massless (triplet) ISF in \eqref{16092025-01-man03-10}, the fields $\phi^{\mu_1\ldots \mu_s}(p)$ with $s=0,1$, and $n\geq 2$ stand for the  respective scalar, vector, and totally symmetric tensor fields of the Lorentz algebra $so(d-1,1)$ algebra.

The fields in \eqref{16092025-01-man03-10} should satisfy divergence-free and traceless constraints. To simplify the presentation of the constraints we use an index-free notation. To this end we introduce oscillators  $\alpha^\mu$, $\zeta$ and collect the set of fields given \eqref{16092025-01-man03-10} into the respective ket-vectors defined as
{\small
\beq
\label{16092025-01-man03-15} && \hspace{-1cm} \phi_s(p,\alpha,\zeta) := \sum_{n=0}^s  \frac{\zeta^{s-n}}{n!\sqrt{(s-n)!}} \alpha^{\mu_1} \ldots \alpha^{\mu_n}
\phi^{\mu_1\ldots \mu_n}(p)\,, \hspace{0.5cm} \hbox{for massive (triplet) ISF};
\nonumber\\
&& \hspace{-1cm} \phi_s(p,\alpha) := \frac{1}{s!} \alpha^{\mu_1} \ldots \alpha^{\mu_s} \phi^{\mu_1\ldots \mu_s}(p) \,, \hspace{3.3cm} \hbox{for massless (triplet) ISF};
\eeq
}
\!where the argument $\alpha$ in ket-vectors $\phi_s(p,\alpha,\zeta)$ and $\phi_s(p,\alpha)$ \eqref{16092025-01-man03-15} is used to denote the  oscillators $\alpha^\mu$. The constraints can then be presented as (for the notation, see appendix \ref{app-notation}),
\beq
\label{16092025-man03-16} && \Gb\phi_s = 0 \,,
\\
\label{16092025-man03-17} && \Gb   : = \left\{
\begin{array}{llll}
p^\mu\alphab^\mu  - m\zetab  \,, & N_\alpha + N_\zeta - s  \,,   &    &  \hbox{for massive triplet ISF};
\\[7pt]
p^\mu\alphab^\mu \,, & N_\alpha -s \,,   &  & \hbox{for massless triplet ISF};
\\[7pt]
p^\mu\alphab^\mu  - m\zetab  \,, & N_\alpha + N_\zeta - s  \,, \quad & \alphab^\mu \alphab^\mu + \zetab^2\,,   &  \hbox{for massive ISF};
\\[7pt]
p^\mu\alphab^\mu  & N_\alpha -s  &  \alphab^\mu \alphab^\mu\,,  & \hbox{for massless ISF}.
\end{array}\right.\qquad
\eeq
Constraint \eqref{16092025-man03-16} corresponding to the operators $\Gb=N_\alpha + N_\zeta - s$ and $\Gb=N_\alpha  - s$ will be referred to as spin-level constraint.

For example, considering the massless ISF, we make a brief clarifying comment on constraints \eqref{16092025-man03-16}. For $\Gb=p^\mu\alphab^\mu$ and $\Gb=\alphab^\mu\alphab^\mu$, the constraint \eqref{16092025-man03-16} is realized as the respective divergence-free constraint and traceless constraint.
For massless ISF, the spin-level constraint \eqref{16092025-man03-16} tells us that the ket-vector of massless ISF \eqref{16092025-01-man03-15} realized as a degree-$s$ homogeneous polynomial in the oscillators $\alpha^\mu$.
From \eqref{16092025-man03-17}, we see that the triplet ISFs are not subjected to the tracelessness constraint. Discussion of various Lorentz covariant formulation of the triplet ISFs may be found, e.g., in refs.\cite{Bengtsson:1986ys,Pashnev:1989gm,Francia:2002pt,Sagnotti:2003qa, Sorokin:2008tf,Sorokin:2018djm,Campoleoni:2012th}.

Constraints \eqref{16092025-man03-16} are invariant under on-shell gauge symmetries. To describe  gauge symmetries we introduce gauge transformation parameters $\lambda^{\mu_1\ldots \mu_n}(p)$, $n=0,1,\ldots s$, for massive fields, and $\lambda^{\mu_1\ldots \mu_s}(p)$, for massless fields. By using the oscillators, the gauge transformation parameters are collected into the corresponding ket-vectors $\lambda_s(p,\alpha,\zeta)$ and $\lambda_s(p,\alpha)$. The ket-vectors $\lambda_s(p,\alpha)$ for massive/massless fields are obtained by making the replacement $\phi\rightarrow \lambda$ in \eqref{16092025-01-man03-15}. The ket-vector $\lambda_s$ satisfies the constraints $\Gb\lambda_s=0$, where the operator $\Gb$ takes the same form as the one in \eqref{16092025-man03-17}. The gauge transformations of massive/massless (triplet) ISF can then be presented as
\be
\label{16092025-01-man03-30} \delta \phi_s  = G \lambda_{s-1}\,,
\qquad G := \left\{
\begin{array}{ll}
p^\mu\alpha^\mu - m\zeta, \qquad & \hbox{for massive (triplet) ISF};
\\[5pt]
p^\mu\alpha^\mu, & \hbox{for massless (triplet) ISF}.\qquad
\end{array}\right.
\ee

\noinbf{Tower of triplet ISFs}. In order to treat the ISFs with different values of spins on an equal footing we find it convenient to use the tower of massive/massless fields and the corresponding tower of gauge transformation parameters given by
\be   \label{16092025-01-man03-40}
\phi = \sum_{s=0}^\infty \phi_s\,, \qquad \lambda = \sum_{s=0}^\infty \lambda_s\,.
\ee
Considering \eqref{16092025-01-man03-40} for the case of triplet massive/massless ISFs $\phi_s$ we get a tower of massive/massless triplet ISFs $\phi$ which obeys simple set of constraints. Namely for such $\phi$, the constraints and gauge transformations take the form.
{\small
\beq
\label{16092025-man03-46} &&  \Gb\phi = 0 \,, \hspace{3cm}  \delta\phi= G\lambda \,
\\
&& \Gb   : = \left\{
\begin{array}{l}
p^\mu\alphab^\mu  - m\zetab  \,,
\\[7pt]
p^\mu\alphab^\mu \,,
\end{array}\right.
\qquad G   : = \left\{
\begin{array}{ll}
p^\mu\alpha^\mu  - m\zeta  \,,  &  \hbox{for tower of massive triplet ISF};
\\[7pt]
p^\mu\alpha^\mu \,,   &  \hbox{for tower of massless triplet ISF}.
\end{array}\right.\qquad
\eeq
}
Considering \eqref{16092025-01-man03-40} for the case of massless ISFs $\phi_s(p,\alpha)$, we get tower of massless ISFs $\phi(p,\alpha)$ which has the same field content as the one in higher-spin field theory in ref.\cite{Vasiliev:1990en}. Recent investigations of higher-spin field theory may be found, e.g., in refs. \cite{Tatarenko:2024csa,Didenko:2025xca,Korybut:2025vdn,Gelfond:2025alv, Bekaert:2025azj,Skvortsov:2025ohi,Serrani:2025owx,Tran:2025uad,Ivanovskiy:2025kok}.

\vspace{-0.2cm}
\section{ \large On-shell cubic action} \label{sec-03}

\noinbf{Formal cubic action}. At cubic order in the fields, manifestly Lorentz invariant on-shell action can be presented as
\be \label{17092025-man03-01}
S_\smp3 = \int d\Gamma_\smp3  \Phi_\smp3^\dagger \cdot \LL_\smp3  \,,
\ee
where we use the notation%
\footnote{For the case when the action \eqref{17092025-man03-01} is not hermitian it should be  supplemented by suitable hermitian conjugated cousin. Note also that our fields are singlets of an internal symmetry algebra. We expect that incorporation of internal symmetries could be done, e.g., as in refs.\cite{Konstein:1989ij,Metsaev:1991nb,Skvortsov:2020wtf}.
}
\beq
\label{17092025-man03-05} && \Phi_\smp3^\dagger = \prod_{a=1,2,3} \phi_a^\dagger(p_a,\xi_a,\alpha_a,\zeta_a)\,,
\nonumber\\
&& d\Gamma_\smp3 = (2\pi)^d \delta^d(\sum_{a=1,2,3} p_a)\prod_{a=1,2,3}d\Gamma_a\,,
\\
\label{17092025-man03-06} && d\Gamma_a  : = \left\{
\begin{array}{ll}
2|m_a|^{4-d} \delta(p_a^2+m_a^2)\delta(p_a\xi_a)\delta(\xi_a^2)d^d p_ad^d \xi_a\,,  &  \hbox{for massive CSF};
\\[7pt]
2\kappa_a^{4-d}\delta(p_a^2)\delta(p_a\xi_a)\delta(\xi_a^2-\kappa_a^2)d^d p_ad^d \xi_a\,, & \hbox{for massless CSF};
\\[7pt]
2\delta(p_a^2+m_a^2)d^d p_a \,,  &  \hbox{for massive ISF};
\\[7pt]
2\delta(p_a^2)d^d p_a\,, & \hbox{for massless ISF};
\end{array}\right.
\eeq
and density $\LL_\smp3$ \eqref{17092025-man03-01} depends on the momenta $p_a^\mu$, the vectors $\xi_a^\mu$, and the oscillators $\alpha_a^\mu$, $\zeta_a$, $a=1,2,3$,
\be \label{17092025-man03-07}
\LL_\smp3 = \LL_\smp3(p_a^\mu,\xi_a^\mu,\alpha_a^\mu,\zeta_a)\,.
\ee
In \eqref{17092025-man03-01}, the dot $\cdot$ stands for the inner product defined by relation \eqref{20082025-man03-15} in appendix \ref{app-notation}.
The density $\LL_\smp3$ subject to equations which are presented below. We will refer the density $\LL_\smp3$ to as cubic vertex.  Note that to respect  constraints for the momenta $p_a^\mu$ and the vectors $\xi_a^\mu$ corresponding to CSF  \eqref{16092025-01-man03-01}, \eqref{16092025-01-man03-02} and the on-shell conditions $p^2+m^2=0$ ($p^2=0$), for massive (massless) fields we inserted the corresponding $\delta$-functions in the integration measure $d\Gamma_\smp3$.%
\footnote{In the expression for $d\Gamma_a$ \eqref{17092025-man03-06} (and in \eqref{16092025-01-man03-01}), we note the symmetry between massive and massless CSFs under the interchange $|m|\leftrightarrow \kappa$, $p^\mu \leftrightarrow \xi^\mu$. The $\delta^d(\sum_{a=1,2,3} p_a^\mu)$ in \eqref{17092025-man03-05} and constraints \eqref{16092025-01-man03-02} break this symmetry.}

Our aim is to find cubic vertex for $\LL_\smp3$ for values of masses and spins shown in \eqref{15092025-man03-01}-\eqref{15092025-man03-19}.
To this end we present the equations for the vertex $\LL_\smp3$,
\be \label{17092025-man03-10}
\Gb_a \LL_\smp3 = 0 \,, \qquad a=1,2,3 \,,
\ee
\be
\label{17092025-man03-16} \Gb_a   : = \left\{
\begin{array}{llll}
\xi_a^\mu \partial_{\xi_a^\mu} - \SSc_a\,,  & & & \hbox{for massive CSF};
\\[7pt]
 p_a^\mu\partial_{\xi_a^\mu} -\irm\,, & & & \hbox{for massless CSF};
\\[7pt]
p_a^\mu\alphab_a^\mu  - m_a\zetab_a  \,, &  \quad &    & \hspace{-1cm} \hbox{for tower of massive triplet ISF};
\\[7pt]
p_a^\mu\alphab_a^\mu \,,    &  &  & \hspace{-1cm}\hbox{for tower of massless triplet ISF};
\\[7pt]
p_a^\mu\alphab_a^\mu  - m_a\zetab_a  \,, & N_{\alpha_a} + N_{\zeta_a} - s_a  \,,   &    &  \hbox{for massive triplet ISF};
\\[7pt]
p_a^\mu\alphab_a^\mu \,, & N_{\alpha_a} -s_a \,,   &  & \hbox{for massless triplet ISF};
\\[7pt]
p_a^\mu\alphab_a^\mu  - m_a\zetab_a  \,, & N_{\alpha_a} + N_{\zeta_a} - s_a  \,, \quad & \alphab_a^\mu \alphab_a^\mu + \zetab_a^2\,,   &  \hbox{for massive ISF};
\\[7pt]
p_a^\mu\alphab_a^\mu  & N_{\alpha_a} -s_a  &  \alphab_a^\mu \alphab_a^\mu\,,  & \hbox{for massless ISF}.
\end{array}\right.\qquad
\ee

Assume that the $\LL_\smp3$ obeys equations \eqref{17092025-man03-10}. We then note that the close inspection of the expression for the action \eqref{17092025-man03-01} shows that the action turns out to be divergent. All divergences are related to CSFs entering the action. Before more detailed discussion of the divergencies let us present our solution for a finite action.

\noinbf{Finite cubic action}. We find the following modification
of the formal action \eqref{17092025-man03-01}%
\footnote{In the expression for $\chi_a$ in \eqref{17092025-man03-21} we use the light-cone splitting $p^\mu=p^+,p^-,p^I$, $\xi^\mu = \xi^+,\xi^-,\xi^I$ and introduce the notation $\beta:=p^+$.}
\beq
\label{17092025-man03-20} && S_\smp3^\chi = \int d\Gamma_\smp3\,  \delta_\CSFsm^\chi\, \Phi_\smp3^\dagger \cdot \LL_\smp3  \,,
\\
\label{17092025-man03-21} && \delta_\CSFsm^\chi := \prod_{a_\csfsm} \delta(\chi_{a_\csfsm})\,,
\qquad
\chi_a = \left\{
\begin{array}{ll}
k_a\frac{\beta_a}{\xi_a^+} - 1\,, \qquad &  \hbox{for massive CSF};
\\[9pt]
\frac{\xi_a^+}{\beta_a} - k_a \,,  & \hbox{for massless CSF};
\end{array}\right.\qquad
\eeq
where, in \eqref{17092025-man03-21}, the product of $\delta$-functions is performed over the values of the label $a_\csfsm$ corresponding to the CSFs. The constants $k_a$ are restricted to be $k_a\ne 0$. Action \eqref{17092025-man03-20} does not depend on $k_a$. Though the expressions for $\chi_a$ are not manifestly Lorentz invariant we verified that the action  $S_\smp3^\chi$ \eqref{17092025-man03-20} maintains the Lorentz invariance (for the proof, see appendix \ref{lor-inv}).

\noinbf{Light-cone frame and light-cone gauge}. Before to proceed with discussion of the divergencies of action \eqref{17092025-man03-01}, we explain the terminology we use below. Use of {\it light-cone frame} implies that we solve constraints for CSF to express relativistic CSF in terms of the constraint-free CSF (see relation \eqref{22092025-man03-01} in appendix \ref{div}). In the {\it light-cone frame}, the ISF remains to be treated in terms of relativistic gauge fields subject to covariant constraints.
Use of {\it light-cone gauge} implies that we use the just defined light-cone frame for CSF, while, for ISF, we impose light-cone gauge condition and solve the constraints to express the relativistic ISF in terms of constraint-free ISF (see relation \eqref{22092025-man03-50} in appendix \ref{div}).

\noinbf{Divergencies of formal action}. In order to show that the formal action \eqref{17092025-man03-01} is divergent we use the light-cone frame. By definition, the vertex $\LL_\smp3$ should obey  basic equations \eqref{17092025-man03-10}. We then find the following expression for the formal action \eqref{17092025-man03-01}:
\be \label{17092025-man03-36}
S_\smp3|_{\,\lcfrm} = \int d\sigma_\CSFsm^{\vphantom{5pt}}\,  dS_\lcfrm\,,
\ee
where we use the notation
\beq
\label{17092025-man03-30} && d\sigma_\CSFsm^{\vphantom{5pt}} := \prod_a d\sigma_a\,,
\qquad
d\sigma_a := \left\{
\begin{array}{ll}
\frac{1}{\xi_a^+} d\xi_a^+\,, \qquad &  \hbox{for massive CSF}
\\[9pt]
\frac{1}{\beta_a} d\xi_a^+  \,,  & \hbox{for massless CSF}
\end{array}\right.\qquad
\\
&& \hspace{3cm}  dS_\lcfrm-\hbox{independent of $\xi_a^+$},
\eeq
while the explicit expression for a new vertex $dS_\lcfrm$ may be found in appendix \ref{proof}. Here, what is important for us is that the vertex $dS_\lcfrm$ does not depend on the variables $\xi_a^+$.  Taking into account then the expression for $d\sigma_\CSFsm^{\vphantom{5pt}}$ given in \eqref{17092025-man03-30}, we see that each {\it massive CSF leads to logarithmic divergence}, while each {\it massless CSF leads to linear divergence}.
We then note that it is insertion of delta-functions $\delta_\CSFsm^\chi$, \eqref{17092025-man03-21} into the formal action \eqref{17092025-man03-01} that leads to the finite action $S_\smp3^\chi$ \eqref{17092025-man03-20}. Namely, using the relation
\be \label{17092025-man03-35}
\int d\sigma_\CSFsm^{\vphantom{5pt}}\, \delta_\CSFsm^\chi = 1\,,
\ee
we find, the Lorentz invariant action \eqref{17092025-man03-20} considered in the light-cone frame,  takes the form
\be  \label{17092025-man03-40}
S_\smp3^\chi\big|_{\,\lcfrm} = \int dS_\lcfrm\,.
\ee

\noinbf{Finite action in light-cone gauge}. Let us use the notation $S_\smp3^\chi|_\lcgrm$ for finite action \eqref{17092025-man03-20} taken to be in light-cone gauge. For the reader convenience we now explain how $S_\smp3^\chi|_\lcgrm$ is related to  off-shell light-cone gauge action.  To this end we note that, for any light-cone gauge theory, the off-shell light-cone gauge cubic action denoted as $S_{\smp3,\lcgrm}^\offshrm$ is expressed in terms of the cubic Hamiltonian denoted as $P_\smp3^-$ in the following way:
\be \label{17092025-man03-45}
S_{\smp3,\lcgrm}^\offshrm = \int dx^+ P_\smp3^-\,,\qquad  P_\smp3^- =   \int d\Gamma_{\smp3,\lcrm}\,  \Phi_{\smp3,\lcgrm}^{\offshrm\,\dagger}\, \cdot p_\smp3^-\,,
\ee
where $\Phi_{\smp3,\lcgrm}^{\offshrm\,\dagger}$ stands for the product of three off-shell light-cone gauge fields, while $p_\smp3^-$ is the light-cone gauge cubic vertex. Let us use the notation $  S_{\smp3,\lcgrm}^\offshrm|_{\rm on-sh}$ for the action $S_{\smp3,\lcgrm}^\offshrm $ evaluated on the solutions of free equations of motion.%
\footnote{The initial data for the solutions are realized as the fields $\phi_\lcfrm$, $\phi_\lcgrm$ given in \eqref{22092025-man03-01}, \eqref{22092025-man03-50}.}
We then note the matching
\be  \label{17092025-man03-50}
S_\smp3^\chi|_\lcgrm = S_{\smp3,\lcgrm}^\offshrm|_{\rm on-shell}\,.
\ee

\noinbf{Parity-even cubic vertices}. As shown in \eqref{17092025-man03-07}, the cubic vertex $\LL_\smp3$ is depending on the following set of variables:
\be   \label{17092025-man03-60}
p_a^\mu\,,\quad  \xi_a^\mu\,,\quad \alpha_a^\mu\,, \quad \zeta_a\,, \qquad a=1,2,3\,.
\ee
The Lorentz $so(d-1,1)$ algebra symmetries
imply that our cubic vertex $\LL_\smp3$ depends on invariants constructed out of the momenta $p_a^\mu$, vectors of superspace $\xi_a^\mu$, oscillators $\alpha_a^\mu$, the metric tensor $\eta^{\mu\nu}$, and the Levi-Civita symbol $\epsilon^{\mu_1\ldots \mu_d}$. Obviously the oscillators $\zeta_a$ are invariants of the Lorentz algebra. In this paper, to avoid too long study, let us ignore invariants that depend  the Levi-Civita symbol.%
\footnote{ For massless ISFs in $\Ro^{4,1}$, the parity-odd light-cone gauge cubic vertices were studied  in section 8.1 in ref.\cite{Metsaev:2005ar}, while, for the ones in $\Ro^{2,1}$, the parity-odd Lorentz covariant vertices were considered in ref.\cite{Kessel:2018ugi}.
}
This is to say that we restrict our attention to cubic vertices depending on the following invariants of the Lorentz symmetries:
\be  \label{17092025-man03-63}
p_a \xi_b, \quad  p_a\alpha_b\,,\quad \xi_{ab}\,, \quad \xi_a\alpha_b\,,  \quad \alpha_{ab}\,, \quad \zeta_a\,,
\ee
where, in \eqref{17092025-man03-63} and throughout this paper, we use the following shortcuts
\be  \label{17092025-man03-64}
p_a\xi_b := p_a^\mu \xi_b^\mu, \quad  p_a\alpha_b:= p_a^\mu\alpha_b^\mu\,,\quad \xi_{ab} := \xi_a^\mu\xi_b^\mu\,, \quad \xi_a\alpha_b: = \xi_a^\mu \alpha_b^\mu\,,  \quad \alpha_{ab}:= \alpha_a^\mu\alpha_b^\mu\,.
\ee
We refer cubic vertices that depend only on variables shown in \eqref{17092025-man03-63} to as parity-even cubic vertices (or simply to as cubic vertices).

\medskip
\noinbf{Local and non-local vertices}. Let us explain some terminology we use in this paper. If a cubic vertex $\LL_\smp3$ is realized as a finite-order polynomial in the momenta $p_a^\mu$, then we refer such vertex  to as local vertex. Otherwise the cubic vertex  will be referred to as non-local cubic vertex. All solutions for our cubic vertices that involves at least one massive or one massless CSF turn out to be non-local. Depending on the way in which the non-locality is realized in cubic vertices, we use the following terminology:

{\small
\beq
\label{30032025-man01-35} && \hspace{-1.5cm} \begin{array}{ll}
e^{W_\Erm},  \hspace{7mm}   W_\Erm = X_{(1)} \,, \hspace{2cm}   & \hbox{exponential non-locality ($E$-non-locality)};
\\[5pt]
X_\PLrm^\NN, \hspace{7mm}  X_\PLrm = X_{(1)}\,, \hspace{0.5cm} \NN \in \Co /\No_0\,,  & \hbox{power-law non-locality ($PL$-non-locality)};
\\[5pt]
e^{W_\EPLrm},  \hspace{3mm}  W_\EPLrm = X_\smone / Y_\smone   \qquad &\hbox{exponential-power-law non-locality ($EPL$-non-locality)};
\end{array}
\nonumber\\
&& \hspace{3cm} X_\smone\,, \ Y_\smone -\hbox{ are degree-1 polynomials in momenta $p_a^\mu$}.
\eeq
}
The explicit expressions for the degree-1 polynomials $X_{(1)}$, $Y_{(1)}$ entering non-localities \eqref{30032025-man01-35}  may be found below in sections \ref{sec-3csf}-\ref{sec-1csf}.

We now make comment about basic equations for cubic vertices given in \eqref{17092025-man03-10}.

\noinbf{i)} The basic equations for cubic vertices provide us a possibility to find straightforwardly all f-solutions for cubic vertices. These equations can also be used for finding d-solutions. However at present time we are not aware self-contained and systematic method for finding all d-solutions. This is to say that investigation of all d-solutions is beyond the scope of the present time.

\noinbf{ii)} The basic equations for cubic vertices involving CSFs and triplet ISFs are first-order differential equations with respect to vectors $\xi_a^\mu$ and the oscillators $\alpha_a^\mu$, $\zeta_a$. Needless to say that such equations can straightforwardly and systematically be solved by applying the well-known method of characteristics. In the interest of the brevity, let us therefore, in our presentation below given, skip the details of the derivation and present only our results for cubic vertices.

\noinbf{iii)} The basic equations for cubic vertices involving ISFs contain divergence-free, traceless, and spin-level constraints, while ones for triplet ISFs contain the divergence-free and spin level constraints. The divergence-free constraint strongly commutes with traceless constraint. Therefore the knowledge of vertices involving triplet ISFs allow us straightforwardly obtain vertices involving the ISFs. All that is needed to this end is to replace the oscillators $\alpha^\mu$, $\zeta$ in vertices involving triplet ISFs by the modified oscillators $\alpha_\Thsm^\mu$, $\zeta_\Thsm$  that respect the divergence-free constraints and given explicitly in \eqref{20082025-man03-05}, \eqref{20082025-man03-06} in appendix \ref{app-notation}.
As side remark, for action \eqref{17092025-man03-20}, the just mentioned replacement is immaterial because the field $\phi^\dagger$ entering action \eqref{17092025-man03-20} subject to the constraint
$\phi^\dagger (\alpha^2 +\zeta^2)=0$ for massive ISF and $\phi^\dagger\alpha^2=0$ for massless ISF.
Needless to say that the spin-level constraint is easy to solve in our approach.

\section{ \large Cubic vertices for three continuous-spin fields}\label{sec-3csf}

This section is devoted to cubic vertices involving three massive/massless CSFs. Our classification in \eqref{15092025-man03-01}-\eqref{15092025-man03-05}, tells us about  the five particular cases. Let us consider these five cases in turn.

\subsection{    Three massive CSFs}

We consider the cubic vertex that involves three massive CSFs:
\beq
\label{01082025-man03-01} && \hspace{-1cm} (m_1,\SSc_1)_\CSFsm\hbox{-}(m_2,\SSc_2)_\CSFsm\hbox{-}(m_3,\SSc_3)_\CSFsm\,, \qquad    m_1^2 < 0\,,\qquad    m_2^2 < 0\,,\qquad m_3^2 < 0 \,,
\nonumber\\
&& \hspace{-1cm} \hbox{\small three massive continuous-spin fields.}\quad
\eeq
The general f-solution for cubic vertex $\LL_\smp3$ is given by
\beq
\label{01082025-man03-02} && \hspace{-1cm} \LL_\smp3 = L_1^{ \SSc_1} L_2^{ \SSc_2} L_3^{ \SSc_3} V(\Qbf_{12}\,, \Qbf_{23}\,, \Qbf_{32})\,,
\\
\label{01082025-man03-03} &&  L_a =  p_{a-1}^\mu\xi_a^\mu\,, \qquad Q_{aa+1}:=\xi_a^\mu\xi_{a+1}^\mu\,,  \quad \Qbf_{aa+1} = \frac{Q_{aa+1} }{L_a L_{a+1} }\,,
\eeq
where we use notation \eqref{17092025-man03-64} and introduce new vertex $V$ which depends on three variables $\Qbf_{12}$, $\Qbf_{23}$, $\Qbf_{31}$  defined in \eqref{01082025-man03-03}. Let us make the following remarks.

\noinbf{i)} The vertex $V$ is not fixed by equations \eqref{17092025-man03-10} and realized therefore as the freedom of our solution for the cubic vertex $\LL_\smp3$. This implies that we have many cubic interaction vertices.

\noinbf{ii)} From  \eqref{01082025-man03-02}, \eqref{01082025-man03-03}, we note that
the dependence of the cubic vertex $\LL_\smp3$ on the momenta $p_a^\mu$ is governed by the variables $L_1$, $L_2$, $L_3$,  and  $\Qbf_{12}$, $\Qbf_{23}$, $\Qbf_{31}$.

\noinbf{iii)} In general, the cubic vertex $\LL_\smp3$ \eqref{01082025-man03-02} is non-polynomial in the momenta $p_a^\mu$ and therefore non-local. Namely, taking into account the expression for $\SSc$ in \eqref{16092025-01-man03-02} and the pre-factor $L_1^{ \SSc_1}L_2^{\SSc_2}L_3^{ \SSc_3}$ \eqref{01082025-man03-02}, we note that the cubic vertex $\LL_\smp3$ \eqref{01082025-man03-02} is realized $PL$-non-local cubic vertex (see \eqref{30032025-man01-35}).

\noinbf{iv)} To demonstrate explicitly the one-to-one correspondence of our covariant results with the light-cone gauge results in ref.\cite{Metsaev:2025qkr} we use notation \eqref{29092025-man03-01}, \eqref{29092025-man03-08} in appendix \ref{connect}. Using then the notation $X|_\lcgrm$ for the light-cone gauge projection of the variables in \eqref{01082025-man03-02}, we note the relations
\be \label{01082025-man03-10}
L_a|_\lcgrm = \ccc_a L_a^u\,, \quad \Qbf_{aa+1}|_\lcgrm = \Qbf_{aa+1}^{uu}\,,
\ee
where the expressions on r.h.s in \eqref{01082025-man03-10} enter the light-cone gauge cubic vertex in (4.2) in ref.\cite{Metsaev:2025qkr}. The factors $c_a$ in \eqref{01082025-man03-10} lead to the factors $E_a$ in \eqref{29092025-man03-15}.

\noinbf{Formally local vertices}. Solution \eqref{01082025-man03-02} can be represented as
{\small
\beq
\label{01082025-man03-05} && \LL_\smp3 = Q_{12}^{ \half(\SSc_1+\SSc_2-\SSc_3)} Q_{23}^{ \half(\SSc_2+\SSc_3-\SSc_2)} Q_{31}^{ \half(\SSc_3+\SSc_1-\SSc_2)}  V_\locrm(\Qbf_{12}^\locrm\,, \Qbf_{23}^\locrm\,, \Qbf_{32}^\locrm )\,,
\nonumber\\
&& \hspace{1cm} \Qbf_{aa+1}^\locrm : = 1/\Qbf_{aa+1}\,,
\eeq
}
\!where we use the notation as in \eqref{01082025-man03-03} and $V_\locrm$ is a new vertex. As seen from \eqref{01082025-man03-03}, \eqref{01082025-man03-05}, the variables $\Qbf_{aa+1}^\locrm$ are polynomial in the momenta $p_a^\mu$. Requiring the new vertex $V_\locrm$ to be polynomial in $\Qbf_{aa+1}^\locrm$, we see that the vertex $\LL_\smp3$ is also polynomial in the momenta and hence can be considered as the local vertex. The reason why we refer vertex \eqref{01082025-man03-05} to as formally local vertex is as follows. Though $V_\locrm$ is local the integration measure \eqref{17092025-man03-05} involves delta-functions $\delta(p_a\xi_a)$ and hence the non-localities, in the framework of our approach, are unavoidable. As side remark we note that formally local cubic vertex is allowed only for three massive CSFs.

\subsection{   Two massive CSFs (non-equal masses) and one massless CSF}

We consider the cubic vertex that involves two massive CSFs and one massless CSF:
\beq
\label{02082025-man03-01} && \hspace{-1.2cm} (m_1,\SSc_1)_\CSFsm\hbox{-}(m_2,\SSc_2)_\CSFsm\hbox{-}(0,\kappa_3)_\CSFsm\,, \qquad     m_1^2<0\,, \qquad m_2^2<0\,, \qquad m_1^2 \ne m_2^2 \,,
\nonumber\\
&& \hspace{-1.2cm} \hbox{\small two massive CSFs with non-equal masses and one massless CSF.}\quad
\eeq
The general f-solution for cubic vertex $\LL_\smp3$ is given by
\beq
\label{02082025-man03-02} &&  \hspace{-1cm} \LL_\smp3 = e^W L_1^{\SSc_1} L_2^{\SSc_2} V(\Qbf_{12},\Qbf_{23},\Qbf_{31})\,,
\\
\label{02082025-man03-03} && L_1  = p_3 \xi_1\,, \qquad L_2 = p_1\xi_2\,,\quad B_3 = p_2 \xi_3\,,
\qquad  W = \frac{2\irm }{m_2^2-m_1^2}  B_3\,,
\nonumber\\
&& Q_{12} = \xi_{12} \,,
\nonumber\\
&& Q_{23}  = \xi_{23} - \frac{2}{m_1^2-m_2^2} L_2 B_3 \,,\qquad Q_{31} = \xi_{31} + \frac{2}{m_1^2-m_2^2} B_3 L_1\,,
\nonumber\\
&& \Qbf_{12}  = \frac{ Q_{12} }{ L_1 L_2 }\,, \qquad
\Qbf_{23} = \frac{ Q_{23} }{L_2}\,,\qquad \Qbf_{31} = \frac{ Q_{31}}{L_1}\,,
\eeq
where we use notation \eqref{17092025-man03-64} and introduce new vertex $V$ which depends on three variables $\Qbf_{12}$, $\Qbf_{23}$, $\Qbf_{31}$  defined in \eqref{02082025-man03-03}. Let us make the following remarks.

\noinbf{i)} The vertex $V$ is not fixed by equations \eqref{17092025-man03-10} and realized therefore as the freedom of our solution for the cubic vertex $\LL_\smp3$. This implies that we have many cubic interaction vertices.

\noinbf{ii)} From  \eqref{02082025-man03-02}, \eqref{02082025-man03-03}, we note that
the dependence of the cubic vertex $\LL_\smp3$ on the momenta $p_a^\mu$ is governed by the variables $L_1$, $L_2$, $W$,  and  $\Qbf_{12}$, $\Qbf_{23}$, $\Qbf_{31}$.

\noinbf{iii)} The cubic vertex $\LL_\smp3$ \eqref{02082025-man03-02} is non-polynomial in the momenta $p_a^\mu$ and therefore non-local. Namely, taking into account the expression for $\SSc$ in \eqref{16092025-01-man03-02} and the pre-factor $e^W L_1^{ \SSc_1}L_2^{\SSc_2}$ \eqref{02082025-man03-02}, we note that the cubic vertex $\LL_\smp3$ \eqref{02082025-man03-02} is realized as $E$- and $PL$-non-local cubic vertex (see \eqref{30032025-man01-35}).

\noinbf{iv)} To demonstrate explicitly the one-to-one correspondence of our covariant results with the light-cone gauge results in ref.\cite{Metsaev:2025qkr} we use notation \eqref{29092025-man03-01}, \eqref{29092025-man03-08} in appendix \ref{connect}. Using then the notation $X|_\lcgrm$ for the light-cone gauge projection of the variables in \eqref{02082025-man03-02}, we note the relations
\beq
\label{02082025-man03-10} && W|_\lcgrm = W + \frac{\irm \xi_3^+}{\beta_3}\,, \qquad L_1|_\lcgrm = \ccc_1 L_1^u\,, \qquad L_2|_\lcgrm = \ccc_2 L_2^u\,,
\nonumber\\
&& \Qbf_{12}|_\lcgrm = \Qbf_{12}^{uu}\,,  \qquad \Qbf_{23}|_\lcgrm = \ccc_3 \Qbf_{23}^{uu}\,, \qquad \Qbf_{31}|_\lcgrm = \ccc_3 \Qbf_{31}^{uu}\,,
\eeq
where the variables on r.h.s in \eqref{02082025-man03-10} enter the light-cone gauge cubic vertex in (4.5) in ref.\cite{Metsaev:2025qkr}. The $\xi_3^+$-term and the factors $\ccc_1$, $\ccc_2$ in \eqref{02082025-man03-10} lead to the factors $E_3$, $E_1$, $E_2$ in \eqref{29092025-man03-15}.

\subsection{   Two massive CSFs (equal masses) and one massless CSF}

Consider the cubic vertex that involves two massive CSFs (equal masses) and one massless CSF:
\beq
\label{03082025-man03-01} && \hspace{-1cm} (m_1,\SSc_1)_\CSFsm\hbox{-}(m_2,\SSc_2)_\CSFsm\hbox{-}(0,\kappa_3)_\CSFsm\,, \qquad     m_1^2=m^2\,, \qquad m_2^2 = m^2\,, \qquad m^2 < 0 \,,
\nonumber\\
&& \hspace{-1cm} \hbox{\small two massive CSFs with equal masses and one massless CSF.}\quad
\eeq
The general f-solution for cubic vertex $\LL_\smp3$ is given by
\beq
\label{03082025-man03-02} &&  \hspace{-1cm}\LL_\smp3 = e^W   L_1^{\SSc_1} L_2^{\SSc_2} V(B_3\,, \Qbf_{12},\Qbf)\,,
\\
\label{03082025-man03-03} && L_1 = p_3 \xi_1\,, \quad L_2 = p_1 \xi_2\,,\quad B_3 =  p_2\xi_3\,,\quad Q_{ab} = \xi_{ab}\,,
\nonumber\\
&& W = \frac{\irm}{2} \Big(\frac{Q_{31}}{L_1} - \frac{Q_{23}}{L_2} \Big)\,, \quad   \Qbf_{12} = \frac{Q_{12}}{L_1 L_2} \,, \quad \Qbf  = \frac{Q_{31}}{L_1} + \frac{Q_{23}}{L_2} \,,
\eeq
where we use notation \eqref{17092025-man03-64} and introduce new vertex $V$ which depends on three variables $B_3$, $\Qbf_{12}$, $\Qbf$  defined in \eqref{03082025-man03-03}. Let us make the following remarks.

\noinbf{i)} The vertex $V$ is not fixed by equations \eqref{17092025-man03-10} and realized therefore as the freedom of our solution for the cubic vertex $\LL_\smp3$. This implies that we have many cubic interaction vertices.

\noinbf{ii)} From  \eqref{03082025-man03-02}, \eqref{03082025-man03-03}, we note that
the dependence of the cubic vertex $\LL_\smp3$ on the momenta $p_a^\mu$ is governed by the variables $L_1$, $L_2$, $W$,  and  $B_3$, $\Qbf_{12}$, $\Qbf$.

\noinbf{iii)} The cubic vertex $\LL_\smp3$ \eqref{03082025-man03-02} is non-polynomial in the momenta $p_a^\mu$ and therefore non-local. Namely, taking into account the expression for $\SSc$ in \eqref{16092025-01-man03-02} and the pre-factor $e^W L_1^{ \SSc_1}L_2^{\SSc_2}$ \eqref{03082025-man03-02}, we note that the cubic vertex $\LL_\smp3$ \eqref{03082025-man03-02} is realized $PL$- and $EPL$ non-local cubic vertex (see \eqref{30032025-man01-35}).

\noinbf{iv)} To demonstrate explicitly the one-to-one correspondence of our covariant results with the light-cone gauge results in ref.\cite{Metsaev:2025qkr} we use notation \eqref{29092025-man03-01}, \eqref{29092025-man03-08} in appendix \ref{connect}. Using then the notation $X|_\lcgrm$ for the light-cone gauge projection of the variables in \eqref{03082025-man03-02}, we note the relations
\beq
\label{03082025-man03-10} && W|_\lcgrm = W + \frac{\irm \xi_3^+}{\beta_3}\,, \qquad L_1|_\lcgrm = \ccc_1 L_1^u\,,
\qquad L_2|_\lcgrm = \ccc_2 L_2^u\,,
\nonumber\\
&&  B_3|_\lcgrm = \ccc_3 B_3^u\,,\qquad \Qbf_{12}|_\lcgrm = \Qbf_{12}^{uu}\,,  \qquad \Qbf|_\lcgrm = \ccc_3 \Qbf^{uu}\,,
\eeq
where the expressions on r.h.s in \eqref{03082025-man03-10} enter the light-cone gauge cubic vertex in (4.8) in ref.\cite{Metsaev:2025qkr}. The $\xi_3^+$-term and the factors $\ccc_1$, $\ccc_2$ in \eqref{03082025-man03-10} lead to the factors $E_3$, $E_1$, $E_2$ in \eqref{29092025-man03-15}.

\subsection{  One massive CSF and two massless CSFs }

We consider the cubic vertex that involves one massive CSF and two massless CSFs:
\beq
\label{04082025-man03-01}  && \hspace{-1cm} (m_1,\SSc_1)_\CSFsm\hbox{-}(0,\kappa_2)_\CSFsm\hbox{-}(0,\kappa_3)_\CSFsm  \,, \qquad     m_1^2 < 0 \,,
\nonumber\\
&& \hspace{-1cm} \hbox{\small one massive CSF and two massless CSFs.}\quad
\eeq
The general f-solution for cubic vertex $\LL_\smp3$ is given by
\beq
\label{04082025-man03-02}  && \hspace{-1cm} \LL_\smp3 = e^W L_1^{\SSc_1} V(\Qbf_{12}\,, Q_{23}\,, \Qbf_{31})\,,
\\
\label{04082025-man03-03}  && L_1  =  p_3\xi_1\,,\quad B_2 = p_1\xi_2\,, \quad B_3 = p_2\xi_3\,,\qquad W =  \frac{2\irm }{m_1^2} (B_2 - B_3\big) \,,
\nonumber\\
&& Q_{12} = \xi_{12} + \frac{2}{m_1^2} L_1 B_2 \,,
\qquad Q_{23}  = \xi_{23} - \frac{2}{m_1^2} B_2 B_3 \,,
\qquad  Q_{31}  = \xi_{31} + \frac{2}{m_1^2} B_3 L_1\,,\qquad
\nonumber\\
&& \Qbf_{12}  = \frac{ Q_{12}  }{L_1 }\,,\qquad \Qbf_{31}  = \frac{ Q_{31}  }{L_1 }\,,
\eeq
where we use notation \eqref{17092025-man03-64} and introduce new vertex $V$ which depends on three variables $\Qbf_{12}$, $Q_{23}$, $\Qbf_{31}$  defined in \eqref{04082025-man03-03}. Let us make the following remarks.

\noinbf{i)} The vertex $V$ is not fixed by equations \eqref{17092025-man03-10} and realized therefore as the freedom of our solution for the cubic vertex $\LL_\smp3$. This implies that we have many cubic interaction vertices.

\noinbf{ii)} From  \eqref{04082025-man03-02}, \eqref{04082025-man03-03}, we note that
the dependence of the cubic vertex $\LL_\smp3$ on the momenta $p_a^\mu$ is governed by the variables $W$, $L_1$,  and  $\Qbf_{12}$, $Q_{23}$, $\Qbf_{31}$.

\noinbf{iii)} The cubic vertex $\LL_\smp3$ \eqref{04082025-man03-02} is non-polynomial in the momenta $p_a^\mu$ and therefore non-local. Namely, taking into account the expression for $\SSc$ in \eqref{16092025-01-man03-02} and the pre-factor $e^W L_1^{ \SSc_1}$ \eqref{04082025-man03-02}, we note that the cubic vertex $\LL_\smp3$ \eqref{04082025-man03-02} is realized as $E$- and $PL$-non-local cubic vertex (see \eqref{30032025-man01-35}).

\noinbf{iv)} To demonstrate explicitly the one-to-one correspondence of our covariant results with the light-cone gauge results in ref.\cite{Metsaev:2025qkr} we use notation \eqref{29092025-man03-01}, \eqref{29092025-man03-08} in appendix \ref{connect}. Using then the notation $X|_\lcgrm$ for the light-cone gauge projection of the variables in \eqref{04082025-man03-02}, we note the relations
\beq
\label{04082025-man03-10} && W|_\lcgrm = W + \frac{\irm \xi_2^+}{\beta_2} + \frac{\irm \xi_3^+}{\beta_3}\,, \qquad L_1|_\lcgrm = \ccc_1 L_1^u\,,
\nonumber\\
&& \Qbf_{12}|_\lcgrm = \ccc_2 \Qbf_{12}^{uu}\,,  \qquad Q_{23}|_\lcgrm = \ccc_2 \ccc_3 Q_{23}^{uu}\,, \qquad \Qbf_{31}|_\lcgrm = \ccc_3 \Qbf_{31}^{uu}\,,
\eeq
where the expressions on r.h.s in \eqref{04082025-man03-10} enter the light-cone gauge cubic vertex in (4.11) in ref.\cite{Metsaev:2025qkr}.
The $\xi_2^+$- and $\xi_3^+$-terms and the factor $\ccc_1$ in \eqref{04082025-man03-10} lead to the factors $E_2$, $E_3$, $E_1$ in \eqref{29092025-man03-15}.

\subsection{  Three massless CSFs }\label{sec-4-5}

We consider the cubic vertex that involves three massless CSFs:
\beq
\label{05082025-man03-01} && \hspace{-1cm} (0,\kappa_1)_\CSFsm\hbox{-}(0,\kappa_2)_\CSFsm\hbox{-}(0,\kappa_3)_\CSFsm\,,
\qquad \hbox{\small three massless CSFs.}\quad
\eeq
Let us formulate two statements.

\noinbf{Statement 1}. For fields in \eqref{05082025-man03-01}, equations for cubic vertex \eqref{17092025-man03-10} do not admit f-solutions.

\noinbf{Statement 2}. For fields in \eqref{05082025-man03-01}, the particular d-solution to equations for cubic vertex \eqref{17092025-man03-10} can be presented as:
{\small
\beq
\label{05082025-man03-02} &&  \hspace{-1cm} \LL_\smp3 = e^W  V(B_1,B_2,B_3, C_{123})\delta(\sum_{a=1,2,3} B_a)\,,
\\
\label{05082025-man03-03} && W =  \irm \sum_{a=1,2,3}\frac{B_a - B_{a+1}}{3B_a B_{a+1}} \xi_{aa+1}\,,\quad  B_a := p_{a-1} \xi_a \,,\qquad C_{123}  = B_1 \xi_{23} + B_2  \xi_{31} + B_3 \xi_{12}\,,\qquad
\eeq
}
\!where we use notation \eqref{17092025-man03-64} and introduce new vertex $V$ which depends formally on the four variables $B_1$, $B_2$, $B_3$, and $C_{123}$  given in \eqref{05082025-man03-03}. The notation $\delta$ is used for  $1$-dimensional Dirac-delta function. We make the following remarks.

\noinbf{i)} The vertex $V$ is not fixed by equations \eqref{17092025-man03-10} and realized therefore as the freedom of our solution for the cubic vertex $\LL_\smp3$. This implies that we have many cubic interaction vertices.

\noinbf{ii)} From  \eqref{05082025-man03-02}, \eqref{05082025-man03-03}, we note that
the dependence of the cubic vertex $\LL_\smp3$ on the momenta $p_a^\mu$ is governed by the variables $B_1$, $B_2$, $B_3$,  and  $C_{123}$.

\noinbf{iii)} The cubic vertex $\LL_\smp3$ \eqref{05082025-man03-02} is non-polynomial in the momenta $p_a^\mu$ and therefore non-local.

\noinbf{iv)} To demonstrate explicitly the one-to-one correspondence of our covariant results with the light-cone gauge results in ref.\cite{Metsaev:2025qkr} we use notation \eqref{29092025-man03-01}, \eqref{29092025-man03-08} in appendix \ref{connect}. Using then the notation $X|_\lcgrm$ for the light-cone gauge projection of the variables in \eqref{05082025-man03-02}, we note the relations
\be \label{05082025-man03-10}
W|_\lcgrm = W + \sum_{a=1,2,3}\frac{\irm \xi_a^+}{\beta_a}\,, \qquad B_a|_\lcgrm = \ccc_a B_a^u\,,\qquad C_{123}|_\lcgrm = \ccc_1\ccc_2\ccc_3 C_{123}^{uuu}\,,
\ee
where the expressions on r.h.s in \eqref{05082025-man03-10} enter the light-cone gauge cubic vertex in (4.14) in ref.\cite{Metsaev:2025qkr}.
The $\xi_1^+$-, $\xi_2^+$- and $\xi_3^+$-terms in \eqref{05082025-man03-10} lead to the factors $E_1$, $E_2$, $E_3$ in \eqref{29092025-man03-15}.

\section{ \large Cubic vertices for two continuous-spin fields and one integer-spin field}\label{sec-2csf}

This section is devoted to cubic vertices that involve two massive/massless CSFs and one massive/massless ISF. Our classification in \eqref{15092025-man03-06}-\eqref{15092025-man03-12}, tells us about  the seven particular cases. Let us consider these particular cases in turn.

\subsection{  \large Two massive CSFs and one massive ISF}\label{sec-5-1}

We consider the cubic vertex that involves two massive CSFs and one massive ISF:
\beq
\label{06082025-man03-01}  && (m_1,\SSc_1)_\CSFsm\hbox{-}(m_2,\SSc_2)_\CSFsm \hbox{-}(m_3,s_3)_\ISFsm \,,\qquad m_1^2 < 0\,, \qquad m_2^2 < 0\,, \qquad m_3^2 > 0\,,
\nonumber\\
&&  \hbox{\small two massive CSFs and one massive ISF.}\quad
\eeq

\noinbf{Two massive CSFs and one tower of massive triplet ISFs}.
Before to discuss cubic vertex for fields shown in \eqref{06082025-man03-01}, we prefer to consider cubic vertex for two CSFs in \eqref{06082025-man03-01} and one tower of mass-$m_3$ triplet ISFs \eqref{16092025-01-man03-40}. The general f-solution for cubic vertex $\LL_\smp3$ is given by
\beq
\label{06082025-man03-02} && \hspace{-1cm} \LL_\smp3 = L_1^{ \SSc_1} L_2^{ \SSc_2} V(L_3\,, \Qbf_{12}\,, \Qbf_{23}\,, \Qbf_{31}\,;\, Q_{33})\,,
\\
\label{06082025-man03-03} &&  L_1 = p_3 \xi_1\,,\quad L_2 = p_1 \xi_2\,, \quad L_3 =  p_2 \alpha_3 + \frac{m_2^2+m_3^2-m_1^2}{2m_3}\zeta_3\,,
\nonumber\\
&& Q_{12}  =  \xi_{12}\,,\qquad Q_{23}  =  \xi_2 \alpha_3 - \frac{\zeta_3}{m_3} L_2\,, \qquad Q_{31}  =  \alpha_3\xi_1 + \frac{\zeta_3}{m_3} L_1\,,
\nonumber\\
&& \Qbf_{12}  = \frac{Q_{12}}{L_1 L_2}\,, \qquad \Qbf_{23}  = \frac{Q_{23}}{L_2}\,,\qquad \Qbf_{31}  = \frac{Q_{31}}{L_1}\,, \qquad Q_{33} = \alpha_{33} + \zeta_3\zeta_3\,,\qquad
\eeq
where we use notation \eqref{17092025-man03-64} and introduce new vertex $V$ which depends on five variables shown as arguments of $V$ in \eqref{06082025-man03-02} and  defined in \eqref{06082025-man03-03}. Let us make the following remarks.

\noinbf{i)} The vertex $V$ is not fixed by equations \eqref{17092025-man03-10} and realized therefore as the freedom of our solution for the cubic vertex $\LL_\smp3$. This implies that we have many cubic interaction vertices.

\noinbf{ii)} From  \eqref{06082025-man03-02}, \eqref{06082025-man03-03}, we note that
the dependence of the cubic vertex $\LL_\smp3$ on the momenta $p_a^\mu$ is governed by the variables $L_1$, $L_2$,  and  $L_3$, $\Qbf_{12}$, $\Qbf_{23}$, $\Qbf_{31}$, $Q_{33}$.

\noinbf{iii)} The cubic vertex $\LL_\smp3$ \eqref{06082025-man03-02} is non-polynomial in the momenta $p_a^\mu$ and therefore non-local. Namely, taking into account the expression for $\SSc$ in \eqref{16092025-01-man03-02} and the pre-factor $L_1^{ \SSc_1}L_2^{\SSc_2}$ \eqref{06082025-man03-02}, we note that the cubic vertex $\LL_\smp3$ \eqref{06082025-man03-02} is realized as $PL$-non-local cubic vertex (see \eqref{30032025-man01-35}).

\noinbf{iv}) The variables $L_3$, $\Qbf_{23}$, $\Qbf_{31}$, and  $Q_{33}$ are realized as linear and quadratical forms in the oscillators $\alpha_3^\mu$, $\zeta_3$. Therefore in order for the cubic vertex $\LL_\smp3$ \eqref{06082025-man03-02} to be sensible, this cubic vertex is required to be expandable in the just mentioned variables.%
\footnote{If function $f=f(x)$ admits the representation $f=\sum_{n\in\No_0} f_n x^n$, then we refer $f$ to as expandable function.}

\noinbf{v)} To demonstrate explicitly the one-to-one correspondence of our covariant results with the light-cone gauge results in ref.\cite{Metsaev:2025qkr} we use notation \eqref{29092025-man03-01}, \eqref{29092025-man03-08} in appendix \ref{connect}. Using then the notation $X|_\lcgrm$ for the light-cone gauge projection of the variables in \eqref{06082025-man03-02}, we note the relations
\beq
\label{06082025-man03-10} && L_1|_\lcgrm = \ccc_1 L_1^u\,, \qquad L_2|_\lcgrm = \ccc_2 L_2^u\,, \qquad L_3|_\lcgrm =  L_3^\alpha\,, \qquad
\nonumber\\
&& \Qbf_{12}|_\lcgrm = \Qbf_{12}^{uu}\,,  \qquad \Qbf_{23}|_\lcgrm = \Qbf_{23}^{u\alpha}\,,  \qquad \Qbf_{31}|_\lcgrm = \Qbf_{31}^{\alpha u}\,,  \qquad Q_{33}|_\lcgrm = Q_{33}^{\alpha\alpha}\,,
\eeq
where the expressions on r.h.s in \eqref{06082025-man03-10} enter the light-cone gauge cubic vertex in (5.2) in ref.\cite{Metsaev:2025qkr}.
The factors $\ccc_1$, $\ccc_2$ in \eqref{06082025-man03-10} lead to the factors $E_1$, $E_2$ in \eqref{29092025-man03-15}.

\noinbf{Two massive CSFs and one massive triplet ISF}. For two CSFs shown in \eqref{06082025-man03-01} and one massless spin-$s_3$ triplet field, the cubic vertex $\LL_\smp3$ can be presented as in \eqref{06082025-man03-02}, \eqref{06082025-man03-03}, where the vertex $V$ is given by
\be \label{06082025-man03-04}
V = L_3^k \Qbf_{23}^{n_1} \Qbf_{31}^{n_2} Q_{33}^l V_{n_1,n_2,l}(\Qbf_{12})\,,\qquad k = s_3 -n_1 - n_2 - 2 l\,,\qquad  k\,, n_1 \,, n_2 \,, l, \in \No_0\,.
\ee
In \eqref{06082025-man03-04}, we introduced new vertex $V_{n_1,n_2,l}$ which depends only on the variable  $\Qbf_{12}$. The integers $n_1$, $n_2$, $l$ subject to the conditions in \eqref{06082025-man03-04} and the new vertex $V_{n_1,n_2,l}$ describe a freedom of our solution for the cubic vertex $\LL_\smp3$.

\noinbf{Two massive CSFs and one massive ISF}. For fields shown in \eqref{06082025-man03-01}, cubic vertex $\LL_\smp3$ can be presented as in \eqref{06082025-man03-02}, \eqref{06082025-man03-03}, where the vertex $V$ is given by
\be \label{06082025-man03-05}
V = L_3^k \Qbf_{23}^{n_1} \Qbf_{31}^{n_2} V_{n_1,n_2}(\Qbf_{12})\big|_\Thsm \,,\qquad k = s_3 -n_1 - n_2\,,\quad k, n_1, n_2 \in \No_0\,.
\ee
In \eqref{06082025-man03-05}, we introduced a new vertex $V_{n_1,n_2}$ that depends only on the variable  $\Qbf_{12}$. The integers $n_1$, $n_2$ subject to the conditions in \eqref{06082025-man03-05} and the new vertex $V_{n_1,n_2}$ describe a freedom of the solution for the cubic vertex $\LL_\smp3$. The notation $|_\Thsm$ indicates that, in \eqref{06082025-man03-05}, we make the replacement $\alpha_3^\mu\rightarrow \alpha_{\Thsm\,3}^\mu$, $\zeta_3\rightarrow \zeta_{\Thsm\,3}$, where the new oscillators are defined in \eqref{20082025-man03-05} in appendix \ref{app-notation}. In the expression for the action \eqref{17092025-man03-20}, such replacement can be ignored (see remark at the end of section \ref{sec-03}).

\subsection{ \large Two massive CSFs (non-equal masses) and one massless ISF}\label{sec-5-2}

We consider the cubic vertex for two massive CSFs (non-equal masses) and one massless ISF:
\beq
\label{07082025-man03-01} && (m_1,\SSc_1)_\CSFsm\hbox{-}(m_2,\SSc_2)_\CSFsm \hbox{-}(0,s_3)_\ISFsm \,, \qquad m_1^2 < 0\,, \qquad m_2^2 < 0 \qquad m_1^2 \ne m_2^2\,,
\nonumber\\
&&  \hbox{\small two massive CSFs with non-equal masses and one massless ISF.}\quad
\eeq

\noinbf{Two massive CSFs and one tower of massless triplet ISFs}. First, we discuss a cubic vertex for two CSFs shown in \eqref{08082025-man03-01} and one tower of massless triplet ISFs. The general f-solution for the cubic vertex $\LL_\smp3$ is given by
\beq
\label{07082025-man03-02} && \hspace{-1cm} \LL_\smp3 = L_1^{ \SSc_1} L_2^{ \SSc_2} V(\Qbf_{12}\,, \Qbf_{23}\,, \Qbf_{31}; q_{33})\,,
\\
\label{07082025-man03-03} &&  L_1  = p_3\xi_1\,, \qquad L_2  =  p_1\xi_2\,,\quad B_3 = p_2 \alpha_3\,,
\nonumber\\
&& Q_{12}   =  \xi_{12}\,,
\nonumber\\
&& Q_{23}   =  \xi_2 \alpha_3 -  \frac{2}{ m_1^2-m_2^2} L_2 B_3\,,\qquad Q_{31}   =  \alpha_3\xi_1 +  \frac{2}{ m_1^2-m_2^2} B_3  L_1 \,,
\nonumber\\
&& \Qbf_{12}   = \frac{Q_{12} }{L_1L_2}\,, \qquad \Qbf_{23}  = \frac{Q_{23}}{L_2}\,,\qquad \Qbf_{31}  = \frac{Q_{31}}{L_1}\,,\qquad q_{33}: = \alpha_{33}\,,
\eeq
where we use notation \eqref{17092025-man03-64} and introduce new vertex $V$ which depends on four variables shown explicitly in \eqref{07082025-man03-02} and defined in \eqref{07082025-man03-03}. Let us make the following remarks.

\noinbf{i)} The vertex $V$ is not fixed by equations \eqref{17092025-man03-10} and realized therefore as the freedom of our solution for the cubic vertex $\LL_\smp3$. This tells us that we have many cubic interaction vertices.

\noinbf{ii)} From  \eqref{07082025-man03-02}, \eqref{07082025-man03-03}, we note that
the dependence of the cubic vertex $\LL_\smp3$ on the momenta $p_a^\mu$ is governed by the variables $L_1$, $L_2$,  and $\Qbf_{12}$, $\Qbf_{23}$, $\Qbf_{31}$.

\noinbf{iii)} The cubic vertex $\LL_\smp3$ \eqref{07082025-man03-02} is non-polynomial in the momenta $p_a^\mu$ and therefore non-local. Namely, taking into account the expression for $\SSc$ in \eqref{16092025-01-man03-02} and the pre-factor $L_1^{ \SSc_1} L_2^{ \SSc_2}$ \eqref{07082025-man03-02}, we note that the cubic vertex $\LL_\smp3$ \eqref{07082025-man03-02} is realized as $PL$-non-local cubic vertex (see \eqref{30032025-man01-35}).

\noinbf{iv}) The variables $\Qbf_{23}$, $\Qbf_{31}$, $q_{33}$ are realized as linear and quadratical forms in the oscillators $\alpha_3^\mu$. Therefore in order for the cubic vertex $\LL_\smp3$ \eqref{07082025-man03-02} to be sensible, this cubic vertex is required to be expandable in the just mentioned variables.

\noinbf{v)} To demonstrate explicitly the one-to-one correspondence of our covariant results with the light-cone gauge results in ref.\cite{Metsaev:2025qkr} we use notation \eqref{29092025-man03-01}, \eqref{29092025-man03-08} in appendix \ref{connect}. Using then the notation $X|_\lcgrm$ for the light-cone gauge projection of the variables in \eqref{07082025-man03-02}, we note the relations
\beq
\label{07082025-man03-10} && L_1|_\lcgrm = \ccc_1 L_1^u\,, \qquad L_2|_\lcgrm = \ccc_2 L_2^u\,, \qquad \Qbf_{12}|_\lcgrm = \Qbf_{12}^{uu}\,,
\nonumber\\
&& \Qbf_{23}|_\lcgrm = \Qbf_{23}^{u\alpha}\,,  \qquad \Qbf_{31}|_\lcgrm = \Qbf_{31}^{\alpha u}\,,  \qquad q_{33}|_\lcgrm = q_{33}^{\alpha\alpha}\,,
\eeq
where the expressions on r.h.s in \eqref{07082025-man03-10} enter the light-cone gauge cubic vertex in (5.12) in ref.\cite{Metsaev:2025qkr}.
The factors $\ccc_1$, $\ccc_2$ in \eqref{07082025-man03-10} lead to the factors $E_1$, $E_2$ in \eqref{29092025-man03-15}.

\noinbf{Two massive CSFs and one massless triplet ISF}. For two CSFs shown in \eqref{07082025-man03-01} and one massless spin-$s_3$ triplet field, the cubic vertex $\LL_\smp3$ can be presented as in \eqref{07082025-man03-02}, \eqref{07082025-man03-03}, where the vertex $V$ is given by
\be \label{07082025-man03-04}
V =  \Qbf_{23}^{n_1} \Qbf_{31}^{n_2} q_{33}^l V_{n_1,n_2,l}(\Qbf_{12})\,, \qquad n_1 + n_2 + 2 l = s_3\,, \qquad \quad n_1\,,n_2\,,l \in \No_0\,.
\ee
In \eqref{07082025-man03-04}, we introduced new vertex $V_{n_1,n_2,l}$ which depends only on the variable  $\Qbf_{12}$. The integers $n_1$, $n_2$, $l$ subject to the conditions in \eqref{07082025-man03-04} and the new vertex $V_{n_1,n_2,l}$ describe a freedom of our solution for the cubic vertex $\LL_\smp3$.

\noinbf{Two massive CSFs and one massless ISF}. For fields shown in \eqref{07082025-man03-01}, cubic vertex $\LL_\smp3$ can be presented as in \eqref{07082025-man03-02}, \eqref{07082025-man03-03}, where the vertex $V$ is given by
\be \label{07082025-man03-05}
V =   \Qbf_{23}^{n_1} \Qbf_{31}^{n_2} \, V_{n_1,n_2}(\Qbf_{12})\big|_\Thsm \,,
\hspace{1cm} n_1 + n_2 = s_3 \,,\qquad n_1\,, n_2 \in \No_0\,.
\ee
In \eqref{07082025-man03-05}, we introduced a new vertex $V_{n_1,n_2}$ that depends only on the variable  $\Qbf_{12}$. The integers $n_1$, $n_2$ subject to the conditions in \eqref{07082025-man03-05} and the new vertex $V_{n_1,n_2}$ describe a freedom of the solution for the cubic vertex $\LL_\smp3$. The notation $|_\Thsm$ indicates that, in \eqref{07082025-man03-05}, we make the replacement $\alpha_3^\mu\rightarrow \alpha_{\Thsm\,3}^\mu$, where the new oscillator $\alpha_\Thsm^\mu$  is given in \eqref{20082025-man03-06} in appendix \ref{app-notation}. In the expression for the action \eqref{17092025-man03-20}, such replacement can be ignored (see our remark at the end of section \ref{sec-03}).

\subsection{ \large Two massive CSFs (equal masses) and one massless ISF}

We consider the cubic vertex for two massive CSFs (equal masses) and one massless ISF:
\beq
\label{08082025-man03-01} && (m_1,\SSc_1)_\CSFsm\hbox{-}(m_2,\SSc_2)_\CSFsm \hbox{-}(0,s_3)_\ISFsm \,, \qquad m_1^2 = m^2\,, \qquad m_2^2 = m^2\,, \qquad m^2 < 0\,,
\nonumber\\
&&  \hbox{\small two massive CSFs with equal masses and one massless ISF.}\quad
\eeq

\noinbf{Two massive CSFs and one tower of massless triplet ISFs}.
First, we discuss a cubic vertex for two CSFs shown in \eqref{08082025-man03-01} and one tower of massless triplet ISFs. The general f-solution for the cubic vertex $\LL_\smp3$ is given by
\beq
\label{08082025-man03-02} && \hspace{-1cm} \LL_\smp3 = L_1^{\SSc_1} L_2^{\SSc_2} V(B_3,\Qbf_{12}\,, \Cbf_{123}; q_{33})\,,
\\
\label{08082025-man03-03} && L_1 =  p_3 \xi_1\,, \quad L_2 = p_1 \xi_2\,, \quad B_3 = p_2 \alpha_3\,,
\nonumber\\
&& Q_{12}  =  \xi_{12}\,, \qquad  C_{123}  =  L_1 \xi_2 \alpha_3 + L_2 \alpha_3 \xi_1 \,,
\nonumber\\
&& \Qbf_{12} = \frac{Q_{12}}{L_1 L_2}\,, \qquad \Cbf_{123}  = \frac{C_{123}}{L_1 L_2}\,, \qquad q_{33}:= \alpha_{33}\,,
\eeq
where we use notation \eqref{17092025-man03-64} and introduce new vertex $V$ which depends on four variables shown explicitly in \eqref{08082025-man03-02} and defined in \eqref{08082025-man03-03}. Let us make the following remarks.

\noinbf{i)} The vertex $V$ is not fixed by equations \eqref{17092025-man03-10} and realized therefore as the freedom of our solution for the cubic vertex $\LL_\smp3$. This tells us that we have many cubic interaction vertices.

\noinbf{ii)} From  \eqref{08082025-man03-02}, \eqref{08082025-man03-03}, we note that
the dependence of the cubic vertex $\LL_\smp3$ on the momenta $p_a^\mu$ is governed by the variables $L_1$, $L_2$,  and $B_3$, $\Qbf_{12}$, $\Cbf_{123}$.

\noinbf{iii)} The cubic vertex $\LL_\smp3$ \eqref{08082025-man03-02} is non-polynomial in the momenta $p_a^\mu$ and therefore non-local. Namely, taking into account the expression for $\SSc$ in \eqref{16092025-01-man03-02} and the pre-factor $L_1^{ \SSc_1} L_2^{ \SSc_2}$ \eqref{08082025-man03-02}, we note that the cubic vertex $\LL_\smp3$ \eqref{08082025-man03-02} is realized as $PL$-non-local cubic vertex (see \eqref{30032025-man01-35}).

\noinbf{iv}) The variables $B_3$, $\Cbf_{123}$, and $q_{33}$ are realized as linear and quadratical forms in the oscillators $\alpha_3^\mu$. Therefore in order for the cubic vertex $\LL_\smp3$ \eqref{08082025-man03-02} to be sensible, this cubic vertex is required to be expandable in the just mentioned variables.

\noinbf{v)} To demonstrate explicitly the one-to-one correspondence of our covariant results with the light-cone gauge results in ref.\cite{Metsaev:2025qkr} we use notation \eqref{29092025-man03-01}, \eqref{29092025-man03-08} in appendix \ref{connect}. Using then the notation $X|_\lcgrm$ for the light-cone gauge projection of the variables in \eqref{08082025-man03-02}, we note the relations
\beq
\label{08082025-man03-10} && L_1|_\lcgrm = \ccc_1 L_1^u\,, \qquad L_2|_\lcgrm = \ccc_2 L_2^u\,, \qquad B_3|_\lcgrm = B_3^\alpha\,,
\nonumber\\
&& \Qbf_{12}|_\lcgrm = \Qbf_{12}^{uu}\,,  \qquad \Cbf_{123}|_\lcgrm = \Cbf_{123}^{uu\alpha}\,,  \qquad \qquad q_{33}|_\lcgrm = q_{33}^{\alpha\alpha}\,,
\eeq
where the expressions on r.h.s in \eqref{08082025-man03-10} enter the light-cone gauge cubic vertex in (5.17) in ref.\cite{Metsaev:2025qkr}.
The factors $\ccc_1$, $\ccc_2$ in \eqref{08082025-man03-10} lead to the factors $E_1$, $E_2$ in \eqref{29092025-man03-15}.

\noinbf{Two massive CSFs and one massless triplet ISF}. For two CSFs shown in \eqref{08082025-man03-01} and one massless spin-$s_3$ triplet field, the cubic vertex $\LL_\smp3$ can be presented as in \eqref{08082025-man03-02}, \eqref{08082025-man03-03}, where the vertex $V$ is given by
\be \label{08082025-man03-04}
V = B_3^k \Cbf_{123}^n q_{33}^l V_{n,l}(\Qbf_{12})\,, \hspace{1cm}   k = s_3 - n - 2 l\,, \qquad k\,, n \,, l \in \No_0\,.\qquad
\ee
In \eqref{08082025-man03-04}, we introduced new vertex $V_{n,l}$ which depends only on the variable  $\Qbf_{12}$. The integers $n$, $l$ subject to the conditions in \eqref{08082025-man03-04} and the new vertex $V_{n,l}$ describe a freedom of our solution for the cubic vertex $\LL_\smp3$.

\noinbf{Two massive CSFs and one massless ISF}. For fields shown in \eqref{08082025-man03-01}, cubic vertex $\LL_\smp3$ can be presented as in \eqref{08082025-man03-02}, \eqref{08082025-man03-03}, where the vertex $V$ is given by
\be \label{08082025-man03-05}
V =   B_3^k \Cbf_{123}^n \, V_n(\Qbf_{12})\big|_\Thsm \,,\qquad k = s_3 -n\,,\qquad k\,, n \in \No_0 \,.
\ee
In \eqref{08082025-man03-05}, we introduced a new vertex $V_n$ that depends only on the variable  $\Qbf_{12}$. The integer $n$ subject to the conditions in \eqref{08082025-man03-05} and the new vertex $V_n$ describe a freedom of the solution for the cubic vertex $\LL_\smp3$. The notation $|_\Thsm$ indicates that, in \eqref{08082025-man03-05}, we make the replacement $\alpha_3^\mu\rightarrow \alpha_{\Thsm\,3}^\mu$, where the new oscillator $\alpha_\Thsm^\mu$  is given in \eqref{20082025-man03-06} in appendix \ref{app-notation}. In the expression for the action \eqref{17092025-man03-20}, such replacement can be ignored (see our remark at the end of section \ref{sec-03}).

\subsection{ \large One massive CSF, one massless CSF and one massive ISF}

We consider the cubic vertex for one massive CSF, one massless CSF and one massive ISF:
\beq
\label{09082025-man03-01} && (m_1,\SSc_1)_\CSFsm\hbox{-}(0,\kappa_2)_\CSFsm \hbox{-}(m_3,s_3)_\ISFsm \,,\qquad m_1^2 < 0 \,, \qquad m_3^2 > 0 \,,
\nonumber\\
&&  \hbox{\small one massive CSF, one massless CSF, and one massive ISF.}\quad
\eeq
\noinbf{One massive CSF, one massless CSF, and one tower of massive triplet ISFs}.
First, we discuss a cubic vertex for two CSFs shown in \eqref{09082025-man03-01} and one tower of mass-$m_3$ triplet ISFs. The general f-solution for the cubic vertex $\LL_\smp3$ is given by
\beq
\label{09082025-man03-02} &&  \hspace{-1cm} \LL_\smp3 = e^W L_1^{\SSc_1}   V(L_3,\Qbf_{12}\,, Q_{23}\,,\Qbf_{31}; Q_{33})\,,
\\
\label{09082025-man03-03} &&  L_1 =  p_3 \xi_1\,, \quad B_2=p_1\xi_2\,, \quad L_3 = p_2 \alpha_3 + \frac{m_3^2-m_1^2}{2m_3}\zeta_3\,,
\nonumber\\
&& W = \frac{2\irm  }{m_1^2-m_3^2} B_2\,,
\nonumber\\
&& Q_{12}  =  \xi_{12} -  \frac{2}{m_3^2-m_1^2} L_1 B_2  \,,
\nonumber\\
&& Q_{23}  =  \xi_2\alpha_3 - \frac{\zeta_3}{m_3} B_2  +  \frac{2}{m_3^2-m_1^2} B_2  L_3 \,,\qquad  Q_{31}  =  \alpha_3 \xi_1 + \frac{\zeta_3}{m_3}L_1\,,
\nonumber\\
&& \Qbf_{12}  = \frac{Q_{12}}{L_1}\,, \qquad
\Qbf_{31}  = \frac{Q_{31}}{L_1}\,, \hspace{1cm}  Q_{33}= \alpha_{33} + \zeta_3\zeta_3\,,
\eeq
where we use notation \eqref{17092025-man03-64} and introduce new vertex $V$ which depends on five variables shown explicitly in \eqref{09082025-man03-02} and defined in \eqref{09082025-man03-03}. Let us make the following remarks.

\noinbf{i)} The vertex $V$ is not fixed by equations \eqref{17092025-man03-10} and realized therefore as the freedom of our solution for the cubic vertex $\LL_\smp3$. This tells us that we have many cubic interaction vertices.

\noinbf{ii)} From  \eqref{09082025-man03-02}, \eqref{09082025-man03-03}, we note that
the dependence of the cubic vertex $\LL_\smp3$ on the momenta $p_a^\mu$ is governed by the variables $W$, $L_1$,  and $L_3$, $\Qbf_{12}$, $Q_{23}$, $\Qbf_{31}$.

\noinbf{iii)} The cubic vertex $\LL_\smp3$ \eqref{09082025-man03-02} is non-polynomial in the momenta $p_a^\mu$ and therefore non-local. Namely, taking into account the expression for $\SSc$ in \eqref{16092025-01-man03-02} and the pre-factor $e^W L_1^{ \SSc_1}$ \eqref{09082025-man03-02}, we note that the cubic vertex $\LL_\smp3$ \eqref{09082025-man03-02} is realized as $E$- and $PL$-non-local cubic vertex (see \eqref{30032025-man01-35}).

\noinbf{iv}) The variables $L_3$, $Q_{23}$ , $\Qbf_{31}$, and $Q_{33}$ are realized as linear and quadratical forms in the oscillators $\alpha_3^\mu$, $\zeta_3$. Therefore in order for the cubic vertex $\LL_\smp3$ \eqref{09082025-man03-02} to be sensible, this cubic vertex is required to be expandable in the just mentioned variables.

\noinbf{v)} To demonstrate explicitly the one-to-one correspondence of our covariant results with the light-cone gauge results in ref.\cite{Metsaev:2025qkr} we use notation \eqref{29092025-man03-01}, \eqref{29092025-man03-08} in appendix \ref{connect}. Using then the notation $X|_\lcgrm$ for the light-cone gauge projection of the variables in \eqref{09082025-man03-02}, we note the relations
\beq
\label{09082025-man03-10} && W|_\lcgrm = W + \frac{\irm \xi_2^+}{\beta_2}\,,\qquad L_1|_\lcgrm = \ccc_1 L_1^u\,, \qquad L_3|_\lcgrm = L_3^\alpha\,,
\nonumber\\
&& \Qbf_{12}|_\lcgrm = c_2 \Qbf_{12}^{uu}\,,  \qquad Q_{23}|_\lcgrm = \ccc_2Q_{23}^{u\alpha}\,,  \qquad \Qbf_{31}|_\lcgrm = \Qbf_{31}^{\alpha u}\,,  \qquad Q_{33}|_\lcgrm = Q_{33}^{\alpha\alpha}\,,\qquad
\eeq
where the expressions on r.h.s in \eqref{09082025-man03-10} enter the light-cone gauge cubic vertex in (5.22) in ref.\cite{Metsaev:2025qkr}.
The $\xi_2^+$-term and the factor $\ccc_1$ in \eqref{09082025-man03-10} lead to the factors $E_2$,  $E_1$ in \eqref{29092025-man03-15}.

\noinbf{One massive CSF, one massless CSF, and one massive triplet ISF}. For two CSFs shown in \eqref{09082025-man03-01}, and one mass-$m_3$ and spin-$s_3$ triplet field, the cubic vertex $\LL_\smp3$ can be presented as in \eqref{09082025-man03-02}, \eqref{09082025-man03-03}, where the vertex $V$ is given by
\be \label{09082025-man03-04}
V = L_3^k Q_{23}^{n_1} \Qbf_{31}^{n_2} Q_{33}^l V_{n_1,n_2,l}(\Qbf_{12})\,,\quad  k = s_3 -n_1 - n_2 - 2 l\,, \quad k,n_1,n_2,l\in \No_0\,.
\ee
In \eqref{09082025-man03-04}, we introduced new vertex $V_{n_1,n_2,l}$ which depends only on the variable  $\Qbf_{12}$. The three integers $n_1$, $n_2$, $l$ subject to the conditions in \eqref{09082025-man03-04} and the new vertex $V_{n_1,n_2,l}$ describe a freedom of our solution for the cubic vertex $\LL_\smp3$.

\noinbf{One massive CSF, one massless CSF, and one massive ISF}. For fields shown in \eqref{09082025-man03-01}, cubic vertex $\LL_\smp3$ can be presented as in \eqref{09082025-man03-02}, \eqref{09082025-man03-03}, where the vertex $V$ is given by
\be \label{09082025-man03-05}
V =   L_3^k Q_{23}^{n_1} \Qbf_{31}^{n_2} \, V_{n_1,n_2}(\Qbf_{12})\big|_\Thsm \,,
\qquad k = s_3 -n_1 - n_2\,,\quad k\,, n_1 \,, n_2 \in \No_0\,.
\ee
In \eqref{09082025-man03-05}, we introduced a new vertex $V_{n_1,n_2}$ that depends only on the variable  $\Qbf_{12}$. The integers $n_1$, $n_2$ subject to the conditions in \eqref{09082025-man03-05} and the new vertex $V_{n_1,n_2}$ describe a freedom of the solution for the cubic vertex $\LL_\smp3$. The notation $|_\Thsm$ indicates that, in \eqref{09082025-man03-05}, we make the replacement $\alpha_3^\mu\rightarrow \alpha_{\Thsm\,3}^\mu$, $\zeta_3\rightarrow \zeta_{\Thsm\,3}$, where the new oscillators $\alpha_\Thsm^\mu$,  $\zeta_\Thsm$  are given in \eqref{20082025-man03-05} in appendix \ref{app-notation}. In the expression for the action \eqref{17092025-man03-20}, such replacement can be ignored (see our remark at the end of section \ref{sec-03}).

\subsection{ \large One massive CSF, one massless CSF and one massless ISF}

We consider the cubic vertex for one massive CSF, one massless CSF and one massless ISF:
\beq
\label{10082025-man03-01} && (m_1,\SSc_1)_\CSFsm\hbox{-}(0,\kappa_2)_\CSFsm \hbox{-}(0,s_3)_\ISFsm \,, \qquad m_1^2 < 0 \,,
\nonumber\\
&&  \hbox{\small one massive CSF, one massless CSF, and one massless ISF.}\quad
\eeq

\noinbf{One massive CSF, one massless CSF, and one tower of massless triplet ISFs}.
First, we discuss a cubic vertex for two CSFs shown in \eqref{08082025-man03-01} and one tower of massless triplet ISFs. The general f-solution for the cubic vertex $\LL_\smp3$ is given by
\beq
\label{10082025-man03-02} &&\hspace{-1cm}  \LL_\smp3 = e^W  L_1^{\SSc_1}   V(\Qbf_{12}\,,  Q_{23}\,,\Qbf_{31};q_{33})\,,
\\
\label{10082025-man03-03} && L_1 =  p_3 \xi_1\,,\quad B_2= p_1\xi_2\,,\quad B_3 = p_2\alpha_3\,,\quad W = \frac{2\irm }{m_1^2} B_2\,,
\nonumber\\
&& Q_{12}  =  \xi_{12} +   \frac{2}{m_1^2} L_1 B_2 \,,
\qquad  Q_{23}  =  \xi_2\alpha_3 -   \frac{2}{m_1^2} B_2 B_3    \,,
\qquad  Q_{31}  =  \alpha_3 \xi_1 + \frac{2}{m_1^2} B_3 L_1  \,,\qquad
\nonumber\\
&& \Qbf_{12}  = \frac{Q_{12}}{L_1}\,, \qquad
\Qbf_{31}  = \frac{Q_{31}}{L_1}\,, \qquad q_{33} = \alpha_{33}\,,
\eeq
where we use notation \eqref{17092025-man03-64} and introduce new vertex $V$ which depends on four variables shown explicitly in \eqref{10082025-man03-02} and defined in \eqref{10082025-man03-03}. Let us make the following remarks.

\noinbf{i)} The vertex $V$ is not fixed by equations \eqref{17092025-man03-10} and realized therefore as the freedom of our solution for the cubic vertex $\LL_\smp3$. This tells us that we have many cubic interaction vertices.

\noinbf{ii)} From  \eqref{10082025-man03-02}, \eqref{10082025-man03-03}, we note that
the dependence of the cubic vertex $\LL_\smp3$ on the momenta $p_a^\mu$ is governed by the variables $W$, $L_1$,  and $\Qbf_{12}$, $Q_{23}$, $\Qbf_{31}$.

\noinbf{iii)} The cubic vertex $\LL_\smp3$ \eqref{10082025-man03-02} is non-polynomial in the momenta $p_a^\mu$ and therefore non-local. Namely, taking into account the expression for $\SSc$ in \eqref{16092025-01-man03-02} and the pre-factor $L_1^{ \SSc_1} L_2^{ \SSc_2}$ \eqref{10082025-man03-02}, we note that the cubic vertex $\LL_\smp3$ \eqref{10082025-man03-02} is realized as $E$- and $PL$-non-local cubic vertex (see \eqref{30032025-man01-35}).

\noinbf{iv}) The variables $Q_{23}$, $\Qbf_{31}$, and $q_{33}$ are realized as linear and quadratical forms in the oscillators $\alpha_3^\mu$. Therefore in order for the cubic vertex $\LL_\smp3$ \eqref{10082025-man03-02} to be sensible, this cubic vertex is required to be expandable in the just mentioned variables.

\noinbf{v)} To demonstrate explicitly the one-to-one correspondence of our covariant results with the light-cone gauge results in ref.\cite{Metsaev:2025qkr} we use notation \eqref{29092025-man03-01}, \eqref{29092025-man03-08} in appendix \ref{connect}. Using then the notation $X|_\lcgrm$ for the light-cone gauge projection of the variables in \eqref{10082025-man03-02}, we note the relations
\beq
\label{10082025-man03-10} && W|_\lcgrm = W + \frac{\irm \xi_2^+}{\beta_2}\,,\qquad L_1|_\lcgrm = \ccc_1 L_1^u\,,\qquad \Qbf_{12}|_\lcgrm = \ccc_2 \Qbf_{12}^{uu}\,,
\nonumber\\
&&   Q_{23}|_\lcgrm = \ccc_2 Q_{23}^{u\alpha}\,,  \qquad \Qbf_{31}|_\lcgrm = \Qbf_{31}^{\alpha u}\,,  \qquad q_{33}|_\lcgrm = q_{33}^{\alpha\alpha}\,,\qquad
\eeq
where expressions on r.h.s in \eqref{10082025-man03-10} enter the light-cone gauge cubic vertex in (5.27) in ref.\cite{Metsaev:2025qkr}.
The $\xi_2^+$-term and the factor $\ccc_1$ in \eqref{10082025-man03-10} lead to the factors $E_2$,  $E_1$ in \eqref{29092025-man03-15}.

\noinbf{One massive CSF, one massless CSF, and one massless triplet ISF}. For two CSFs shown in \eqref{10082025-man03-01} and one massless spin-$s_3$ triplet field, the cubic vertex $\LL_\smp3$ can be presented as in \eqref{10082025-man03-02}, \eqref{10082025-man03-03}, where the vertex $V$ is given by
\be \label{10082025-man03-04}
V =  Q_{23}^{n_1} \Qbf_{31}^{n_2} q_{33}^l V_{n_1,n_2,l}(\Qbf_{12})\,,\qquad n_1 + n_2 + 2 l = s_3 \,,   \quad n_1\,, n_2 \,,l \in \No_0\,.
\ee
In \eqref{10082025-man03-04}, we introduced new vertex $V_{n_1,n_2,l}$ which depends only on the variable  $\Qbf_{12}$. The integers $n_1$, $n_2$, $l$ subject to the conditions in \eqref{10082025-man03-04} and the new vertex $V_{n_1,n_2,l}$ describe a freedom of our solution for the cubic vertex $\LL_\smp3$.

\noinbf{One massive CSF, one massless CSF, and one massless ISF}. For fields shown in \eqref{10082025-man03-01}, cubic vertex $\LL_\smp3$ can be presented as in \eqref{10082025-man03-02}, \eqref{10082025-man03-03}, where the vertex $V$ is given by
\be \label{10082025-man03-05}
V =   Q_{23}^{n_1} \Qbf_{31}^{n_2} \, V_{n_1,n_2}(\Qbf_{12})\big|_\Thsm \,, \qquad  n_1 + n_2 = s_3\,, \quad n_1\,, n_2 \in \No_0\,.\qquad
\ee
In \eqref{10082025-man03-05}, we introduced a new vertex $V_{n_1,n_2}$ that depends only on the variable  $\Qbf_{12}$. The integers $n_1$, $n_2$ subject to the conditions in \eqref{10082025-man03-05} and the new vertex $V_{n_1,n_2}$ describe a freedom of the solution for the cubic vertex $\LL_\smp3$. The notation $|_\Thsm$ indicates that, in \eqref{10082025-man03-05}, we make the replacement $\alpha_3^\mu\rightarrow \alpha_{\Thsm\,3}^\mu$, where the new oscillator $\alpha_\Thsm^\mu$  is given in \eqref{20082025-man03-06} in appendix \ref{app-notation}. In the expression for the action \eqref{17092025-man03-20}, such replacement can be ignored (see our remark at the end of section \ref{sec-03}).

\subsection{  Two massless CSFs and one massive ISF}

We consider the cubic vertex for two massless CSFs and one massive ISF:
\beq
\label{11082025-man03-01} && (0,\kappa_1)_\CSFsm\hbox{-}(0,\kappa_2)_\CSFsm \hbox{-}(m_3,s_3)_\ISFsm \,, \qquad m_3^2 > 0\,,
\nonumber\\
&&  \hbox{\small two massless CSFs and one massive ISF.}\quad
\eeq
\noinbf{Two massless CSFs and tower of massive triplet ISFs}.
First, we discuss a cubic vertex for two CSFs shown in \eqref{11082025-man03-01} and one tower of mass-$m_3$ triplet ISFs. The general f-solution for the cubic vertex $\LL_\smp3$ is given by
\beq
\label{11082025-man03-02} && \hspace{-1cm} \LL_\smp3 = e^W  V(L_3\,,  Q_{12}\,,  Q_{23}\,,Q_{31}; Q_{33})\,,
\\
\label{11082025-man03-03} && B_1= p_3\xi_1\,, \quad B_2 = p_1\xi_2\,\quad  L_3 =  p_2 \alpha_3 + \half m_3\zeta_3\,, \qquad  W =  \frac{2\irm}{m_3^2} (B_1 - B_2) \,,\qquad
\nonumber\\
&& Q_{12}  =  \xi_{12} -  \frac{2}{m_3^2} B_1 B_2\,, \hspace{2.5cm} Q_{23}  =  \xi_2 \alpha_3 - \frac{\zeta_3}{m_3} B_2  + \frac{2}{m_3^2} B_2 L_3 \,,
\nonumber\\
&& Q_{31}  =  \alpha_3\xi_1  + \frac{\zeta_3}{m_3} B_1 +  \frac{2}{m_3^2} L_3 B_1 \,, \hspace{1cm} Q_{33} = \alpha_{33} + \zeta_3\zeta_3\,,
\eeq
where we use notation \eqref{17092025-man03-64} and introduce new vertex $V$ which depends on five variables shown explicitly in \eqref{11082025-man03-02} and defined in \eqref{11082025-man03-03}. Let us make the following remarks.

\noinbf{i)} The vertex $V$ is not fixed by equations \eqref{17092025-man03-10} and realized therefore as the freedom of our solution for the cubic vertex $\LL_\smp3$. This tells us that we have many cubic interaction vertices.

\noinbf{ii)} From  \eqref{11082025-man03-02}, \eqref{11082025-man03-03}, we note that
the dependence of the cubic vertex $\LL_\smp3$ on the momenta $p_a^\mu$ is governed by the variables $W$  and $L_3$, $Q_{12}$, $Q_{23}$, $Q_{31}$.

\noinbf{iii)} The cubic vertex $\LL_\smp3$ \eqref{11082025-man03-02} is non-polynomial in the momenta $p_a^\mu$ and therefore non-local. Namely, taking into account the pre-factor $e^W$ \eqref{11082025-man03-02}, we note that the cubic vertex $\LL_\smp3$ \eqref{11082025-man03-02} is realized as $E$-non-local cubic vertex (see \eqref{30032025-man01-35}).

\noinbf{iv}) The variables $L_3$, $Q_{23}$, $Q_{31}$, and $Q_{33}$ are realized as linear and quadratical forms in the oscillators $\alpha_3^\mu$, $\zeta_3$. Therefore in order for the cubic vertex $\LL_\smp3$ \eqref{11082025-man03-02} to be sensible, this cubic vertex is required to be expandable in the just mentioned variables.

\noinbf{v)} To demonstrate explicitly the one-to-one correspondence of our covariant results with the light-cone gauge results in ref.\cite{Metsaev:2025qkr} we use notation \eqref{29092025-man03-01}, \eqref{29092025-man03-08} in appendix \ref{connect}. Using then the notation $X|_\lcgrm$ for the light-cone gauge projection of the variables in \eqref{11082025-man03-02}, we note the relations
\beq
\label{11082025-man03-10} && W|_\lcgrm = W + \frac{\irm \xi_1^+}{\beta_1} + \frac{\irm \xi_2^+}{\beta_2}\,,\qquad  L_3|_\lcgrm = L_3^\alpha\,,\qquad Q_{12}|_\lcgrm = \ccc_1 \ccc_2 Q_{12}^{uu}\,,
\nonumber\\
&&  Q_{23}|_\lcgrm = \ccc_2Q_{23}^{u\alpha}\,,  \qquad Q_{31}|_\lcgrm = \ccc_1 Q_{31}^{\alpha u}\,,  \qquad Q_{33}|_\lcgrm = Q_{33}^{\alpha\alpha}\,,\qquad
\eeq
where expressions on r.h.s in \eqref{11082025-man03-10} enter the light-cone gauge cubic vertex in (5.32) in ref.\cite{Metsaev:2025qkr}.
The $\xi_1^+$- $\xi_2^+$-terms in \eqref{11082025-man03-10} lead to the factors $E_1$,  $E_2$ in \eqref{29092025-man03-15}.

\noinbf{Two massless CSFs and one massive triplet ISF}. For two CSFs shown in \eqref{11082025-man03-01}, and one mass-$m_3$ and spin-$s_3$ triplet field, the cubic vertex $\LL_\smp3$ can be presented as in \eqref{11082025-man03-02}, \eqref{11082025-man03-03}, where the vertex $V$ is given by
\be \label{11082025-man03-04}
V = L_3^k Q_{23}^{n_1} Q_{31}^{n_2} Q_{33}^l V_{n_1,n_2,l}(Q_{12})\,,
\qquad  k = s_3 -n_1 - n_2 - 2 l\,, \qquad k\,, n_1\,, n_2\,, l\in \No_0\,.
\ee
In \eqref{11082025-man03-04}, we introduced new vertex $V_{n_1,n_2,l}$ which depends only on the variable  $\Qbf_{12}$. The integers $n_1$, $n_2$ $l$ subject to the conditions in \eqref{11082025-man03-04} and the new vertex $V_{n_1,n_2,l}$ describe a freedom of our solution for the cubic vertex $\LL_\smp3$.

\noinbf{Two massless CSFs and one massive ISF}. For fields shown in \eqref{11082025-man03-01}, cubic vertex $\LL_\smp3$ can be presented as in \eqref{11082025-man03-02}, \eqref{11082025-man03-03}, where the vertex $V$ is given by
\be \label{11082025-man03-05}
V =   L_3^k Q_{23}^{n_1} Q_{31}^{n_2} \, V_{n_1,n_2}(Q_{12})\big|_\Thsm \,, \qquad k = s_3 -n_1 - n_2\,,\quad k\,,n_1 \,,n_2  \in \No_0\,.
\ee
In \eqref{11082025-man03-05}, we introduced a new vertex $V_{n_1,n_2}$ that depends only on the variable  $\Qbf_{12}$. The integers $n_1$, $n_2$ subject to the conditions in \eqref{11082025-man03-05} and the new vertex $V_{n_1,n_2}$ describe a freedom of the solution for the cubic vertex $\LL_\smp3$. The notation $|_\Thsm$ indicates that, in \eqref{11082025-man03-05}, we make the replacement $\alpha_3^\mu\rightarrow \alpha_{\Thsm\,3}^\mu$, $\zeta_3\rightarrow \zeta_{\Thsm\,3}$, where the new oscillators $\alpha_\Thsm^\mu$, $\zeta_\Thsm$  are given in \eqref{20082025-man03-06} in appendix \ref{app-notation}. In the expression for the action \eqref{17092025-man03-20}, such replacement can be ignored (see our remark at the end of section \ref{sec-03}).

\subsection{  Two massless CSFs and one massless ISF }\label{sub-sec-5-7}

We consider the cubic vertex for two massless CSFs and one massless ISF:
\beq
\label{12082025-man03-01} && (0,\kappa_1)_\CSFsm\hbox{-}(0,\kappa_2)_\CSFsm \hbox{-}(0,s_3)_\ISFsm \,,
\nonumber\\
&&  \hbox{\small two massless CSFs and one massless ISF.}\quad
\eeq

\noinbf{Two massless CSFs and one tower of massless triplet ISFs}.
First, we discuss a cubic vertex for two CSFs shown in \eqref{12082025-man03-01} and one tower of massless triplet ISFs. Let us formulate two statements.

\noinbf{Statement 1}. Equations for cubic interaction vertex \eqref{17092025-man03-10} do not admit f-solutions.

\noinbf{Statement 2}. The particular d-solution to equations for cubic interaction vertex \eqref{17092025-man03-10} can be presented as:
\beq
\label{12082025-man03-02}  && \hspace{-1cm} \LL_\smp3 =  e^W V(B_1,B_2,B_3, C_{123};q_{33})\delta(  B_1  +  B_2)\,,
\\
&& B_1 = p_3 \xi_1\,, \quad B_2 = p_1 \xi_2\,, \quad B_3 = p_2 \alpha_3
\nonumber\\
\label{12082025-man03-03}  && W = \irm \frac{ B_1 - B_2 }{ 2B_1 B_2 } \xi_{12}\,,
\qquad  C_{123} = B_1 \xi_2\alpha_3 + B_2 \alpha_3\xi_1 + B_3 \xi_{12}\,,
\eeq
where we use notation \eqref{17092025-man03-64} and introduce new vertex $V$ which depends formally on the four variables $B_1$, $B_2$, $B_3$, and $C_{123}$  given in \eqref{05082025-man03-03}. The notation $\delta$ is used for  $1$-dimensional Dirac-delta function. We make the following remarks.

\noinbf{i)} The vertex $V$ is not fixed by equations \eqref{17092025-man03-10} and realized therefore as the freedom of our solution for the cubic vertex $\LL_\smp3$. This implies that we have many cubic interaction vertices.

\noinbf{ii)} From  \eqref{12082025-man03-02}, \eqref{12082025-man03-03}, we note that
the dependence of the cubic vertex $\LL_\smp3$ on the momenta $p_a^\mu$ is governed by the variables $B_1$, $B_2$, $B_3$,  and  $C_{123}$.

\noinbf{iii)} The cubic vertex $\LL_\smp3$ \eqref{12082025-man03-02} is non-polynomial in the momenta $p_a^\mu$ and therefore non-local.

\noinbf{iv}) The variables $B_3$, $C_{123}$, $q_{33}$ are realized as linear and quadratical forms in the oscillators $\alpha_3^\mu$. Therefore in order for the cubic vertex $\LL_\smp3$ \eqref{12082025-man03-02} to be sensible, this cubic vertex is required to be expandable in the just mentioned variables.

\noinbf{v)} To demonstrate explicitly the one-to-one correspondence of our covariant results with the light-cone gauge results in ref.\cite{Metsaev:2025qkr} we use notation \eqref{29092025-man03-01}, \eqref{29092025-man03-08} in appendix \ref{connect}. Using then the notation $X|_\lcgrm$ for the light-cone gauge projection of the variables in \eqref{12082025-man03-02}, we note the relations
\beq
\label{12082025-man03-10} && W|_\lcgrm = W + \frac{\irm \xi_1^+}{\beta_1} + \frac{\irm \xi_2^+}{\beta_2}\,,\qquad  B_1|_\lcgrm = \ccc_1 B_1^u\,,\qquad  B_2|_\lcgrm = \ccc_2 B_2^u\,,\qquad  B_3|_\lcgrm = B_3^\alpha\,,
\nonumber\\
&& C_{123}|_\lcgrm = \ccc_1 \ccc_2 C_{123}^{uu\alpha}\,,    \qquad q_{33}|_\lcgrm = q_{33}^{\alpha\alpha}\,,\qquad
\eeq
where expressions on r.h.s in \eqref{12082025-man03-10} enter the light-cone gauge cubic vertex in (5.37) in ref.\cite{Metsaev:2025qkr}.
The $\xi_1$- $\xi_2^+$-terms in \eqref{12082025-man03-10} lead to the factors $E_1$,  $E_2$ in \eqref{29092025-man03-15}.

\noinbf{Two massless CSFs and one massless triplet ISF. d-solution}. For two CSFs shown in \eqref{08082025-man03-01} and one massless spin-$s_3$ triplet field, the d-solution for cubic vertex $\LL_\smp3$ can be presented as in \eqref{12082025-man03-02}, \eqref{12082025-man03-03}, where the vertex $V$ is given by
\be
\label{12082025-man03-04}  V = B_3^k C_{123}^n  q_{33}^l V_{n,l}(B_1,B_2)\,,\qquad
k = s_3 - n - 2 l\,, \quad k\,, n \,, l \in \No_0\,.
\ee
In \eqref{12082025-man03-04}, we introduced new vertex $V_{n,l}$ which depends only on the variables  $B_1$, $B_2$. The two integers $n$, $l$ subject to the conditions in \eqref{12082025-man03-04} and the new vertex $V_{n,l}$ describe a freedom of our solution for the cubic vertex $\LL_\smp3$.

\noinbf{Two massless CSFs and one massless ISF. d-solution}. For fields shown in \eqref{12082025-man03-01}, the d-solution for cubic vertex $\LL_\smp3$ can be presented as in \eqref{12082025-man03-02}, \eqref{12082025-man03-03}, where the vertex $V$ is given by
\be \label{12082025-man03-05}
V = B_3^k C_{123}^n   V_n(B_1,B_2)\big|_\Thsm\,, \qquad   k = s_3 - n\,, \qquad k\,, n \in \No_0\,,
\ee
In \eqref{12082025-man03-05}, we introduced a new vertex $V_n$ that depends only on the variables  $B_1$, $B_2$. The integer $n$ subject to the conditions in \eqref{12082025-man03-05} and the new vertex $V_n$ describe a freedom of the solution for the cubic vertex $\LL_\smp3$. The notation $|_\Thsm$ indicates that, in \eqref{12082025-man03-05}, we make the replacement $\alpha_3^\mu\rightarrow \alpha_{\Thsm\,3}^\mu$, where the new oscillator $\alpha_\Thsm^\mu$  is given in \eqref{20082025-man03-06} in appendix \ref{app-notation}. In the expression for the action \eqref{17092025-man03-20}, such replacement can be ignored (see our remark at the end of section \ref{sec-03}).

\section{ \large Cubic vertices for one continuous-spin fields and two integer-spin fields}\label{sec-1csf}

This section is devoted to cubic vertices that involve one massive/massless CSFs and two massive/massless integers-spin field. Our classification in \eqref{15092025-man03-13}-\eqref{15092025-man03-19}, tells us about  the seven particular cases. Let us consider these particular cases in turn.

\subsection{ \large One massive CSF and two massive ISFs}

We consider the cubic vertex for one massive CSF and two massive ISFs:
\beq
\label{13082025-man03-01} && (m_1,\SSc_1)_\CSFsm\hbox{-}(m_2,s_2)_\ISFsm \hbox{-}(m_3,s_3)_\ISFsm \,,\qquad m_1^2 < 0\,, \qquad m_2^2 > 0 \,, \qquad m_3^2 > 0\,,
\nonumber\\
&&  \hbox{\small one massive CSF and two massive ISFs.}\quad
\eeq
\noinbf{One massive CSF and two towers of massive triplet ISFs}.
First, we discuss a cubic vertex for one CSF shown in \eqref{13082025-man03-01} and two towers of massive  triplet ISFs having masses $m_2$ and $m_3$. The general f-solution for the cubic vertex $\LL_\smp3$ is given by
\beq
\label{13082025-man03-02} && \hspace{-1cm} \LL_\smp3 =  L_1^{\SSc_1} V(L_2\,, L_3\,, \Qbf_{12}\,, Q_{23}\,, \Qbf_{31}; Q_{22}, Q_{33})\,,
\\
\label{13082025-man03-03} &&  L_1 = p_3 \xi_1\,, \quad L_2 =  p_1 \alpha_2 + \frac{m_1^2+m_2^2-m_3^3}{2m_2}\zeta_2\,, \quad L_3 =  p_2 \alpha_3 + \frac{m_2^2+m_3^2-m_1^2}{2m_3}\zeta_3\,,
\nonumber\\
&& Q_{12}   =  \xi_1\alpha_2 - \frac{\zeta_2}{m_2}L_1\,,
\nonumber\\
&& Q_{23}  =   \alpha_{23} + \frac{\zeta_2}{m_2} L_3 - \frac{\zeta_3}{m_3} L_2 + \frac{m_1^2 - m_2^2 - m_3^2}{2m_2m_3}\zeta_2\zeta_3\,,
\nonumber\\
&& Q_{31}  =  \alpha_3\xi_1 + \frac{\zeta_3}{m_3}L_1\,,
\nonumber\\
&& \Qbf_{12}  = \frac{Q_{12}}{L_1}\,, \quad \Qbf_{31}  = \frac{Q_{31}}{L_1}\,,\quad
Q_{aa} = \alpha_{aa} + \zeta_a\zeta_a\,, \quad a=1,2\,,
\eeq
where we use notation \eqref{17092025-man03-64} and introduce new vertex $V$ which depends on seven variables shown explicitly in \eqref{13082025-man03-02} and defined in \eqref{13082025-man03-03}. Let us make the following remarks.

\noinbf{i)} The vertex $V$ is not fixed by equations \eqref{17092025-man03-10} and realized therefore as the freedom of our solution for the cubic vertex $\LL_\smp3$. This tells us that we have many cubic interaction vertices.

\noinbf{ii)} From  \eqref{13082025-man03-02}, \eqref{13082025-man03-03}, we note that
the dependence of the cubic vertex $\LL_\smp3$ on the momenta $p_a^\mu$ is governed by the variables $L_1$ and $L_2$, $L_3$, $\Qbf_{12}$, $Q_{23}$, $\Qbf_{31}$.

\noinbf{iii)} The cubic vertex $\LL_\smp3$ \eqref{13082025-man03-02} is non-polynomial in the momenta $p_a^\mu$ and therefore non-local. Namely, taking into account the expression for $\SSc$ in \eqref{16092025-01-man03-02} and the pre-factor $L_1^{ \SSc_1}$ \eqref{08082025-man03-02}, we note that the cubic vertex $\LL_\smp3$ \eqref{13082025-man03-02} is realized as $PL$-non-local cubic vertex (see \eqref{30032025-man01-35}).

\noinbf{iv}) The variables $L_2$, $L_3$, $\Qbf_{12}$, $Q_{23}$, $\Qbf_{31}$, $Q_{22}$  $Q_{33}$ are realized as linear and quadratical  forms in the oscillators $\alpha_a^\mu$, $\zeta_a$, $a=1,2$.  Therefore in order for the cubic vertex $\LL_\smp3$ \eqref{08082025-man03-02} to be sensible, this cubic vertex is required to be expandable in the just mentioned variables.

\noinbf{v)} To demonstrate explicitly the one-to-one correspondence of our covariant results with the light-cone gauge results in ref.\cite{Metsaev:2025qkr} we use notation \eqref{29092025-man03-01}, \eqref{29092025-man03-08} in appendix \ref{connect}. Using then the notation $X|_\lcgrm$ for the light-cone gauge projection of the variables in \eqref{13082025-man03-02}, we note the relations
\beq
\label{13082025-man03-10} && L_1|_\lcgrm = \ccc_1 L_1^u\,,  \qquad L_2|_\lcgrm =  L_2^\alpha\,, \qquad  L_3|_\lcgrm = L_3^\alpha\,,\qquad   \Qbf_{12}|_\lcgrm = \Qbf_{12}^{u\alpha}\,,
\nonumber\\
&& Q_{23}|_\lcgrm =  Q_{23}^{\alpha\alpha}\,,  \qquad \Qbf_{31}|_\lcgrm = \Qbf_{31}^{\alpha u}\,,  \qquad Q_{22}|_\lcgrm = Q_{22}^{\alpha\alpha}\,,
\qquad Q_{33}|_\lcgrm = Q_{33}^{\alpha\alpha}\,,\qquad
\eeq
where expressions on r.h.s in \eqref{13082025-man03-10} enter the light-cone gauge cubic vertex in (6.2) in ref.\cite{Metsaev:2025qkr}.
The factor $\ccc_1$ in \eqref{13082025-man03-10} leads to the factor $E_1$ in \eqref{29092025-man03-15}.

\noinbf{One massive CSF and two massive triplet ISFs}. For one CSFs shown in \eqref{13082025-man03-01}, one mass-$m_2$ and spin-$s_2$ triplet field, and one mass-$m_3$ and spin-$s_3$ triplet field, the cubic vertex $\LL_\smp3$ can be presented as in \eqref{13082025-man03-02}, \eqref{13082025-man03-03}, where the vertex $V$ is given by
{\small
\beq
\label{13082025-man03-04} && \hspace{-1.2cm} V = L_2^{k_2} L_3^{k_3} \Qbf_{12}^{n_3} Q_{23}^{n_1} \Qbf_{31}^{n_2} Q_{22}^{l_2} Q_{33}^{l_3}\,,
\nonumber\\
&& \hspace{-0.3cm}   k_2 = s_2 - n_1 - n_3 - 2 l_2\,, \quad k_3  = s_3 - n_1 - n_2 - 2 l_3\,,\quad k_2\,, k_3\,, n_1\,, n_2\,, n_3\,, l_2\,, l_3 \in \No_0\,.\qquad
\eeq
}
\!The five integers $n_1$, $n_2$, $n_3$, $l_2$, $l_3$ subject to the conditions in \eqref{13082025-man03-04} describe a freedom of our solution for the cubic vertex $\LL_\smp3$.

\noinbf{One massive CSF and two massive ISFs}. For fields shown in \eqref{13082025-man03-01}, cubic vertex $\LL_\smp3$ can be presented as in \eqref{13082025-man03-02}, \eqref{13082025-man03-03}, where the vertex $V$ is given by
{\small
\beq
\label{13082025-man03-05} && \hspace{-1.2cm} V = L_2^{k_2} L_3^{k_3} \Qbf_{12}^{n_3} Q_{23}^{n_1} \Qbf_{31}^{n_2}\big|_\Thsm\,,
\nonumber\\
&& \hspace{-0.3cm}   k_2 = s_2 - n_1 - n_3 \,, \quad k_3  = s_3 - n_1 - n_2\,,\quad k_2\,, k_3\,, n_1\,, n_2\,, n_3 \in \No_0\,.
\eeq
}
The three integers $n_1$, $n_2$, $n_3$ subject to the conditions in \eqref{13082025-man03-05} describe a freedom of the solution for the cubic vertex $\LL_\smp3$. The notation $|_\Thsm$ indicates that, in \eqref{13082025-man03-05}, we make the replacement $\alpha_a^\mu\rightarrow \alpha_{\Thsm\,a}^\mu$, $\zeta_a\rightarrow \zeta_{\Thsm\,a}$, $a=1,2$, where the new oscillators $\alpha_\Thsm^\mu$, $\zeta_\Thsm$  are given in \eqref{20082025-man03-05} in appendix \ref{app-notation}. In the expression for the action \eqref{17092025-man03-20}, such replacement can be ignored (see our remark at the end of section \ref{sec-03}).

\subsection{  \large One massive CSF, one massive ISF and one massless ISF}

We consider the cubic vertex for one massive CSFs, one massive ISF, and one massless ISF:
\beq
\label{14082025-man03-01} && (m_1,\SSc_1)_\CSFsm\hbox{-}(m_2,s_2)_\ISFsm \hbox{-}(0,s_3)_\ISFsm \,,\qquad m_1^2 < 0\,, \qquad m_2^2 > 0\,,
\nonumber\\
&&  \hbox{\small one massive CSF and one massive ISF, and one massless ISF.}\quad
\eeq
\noinbf{One massive CSF, one tower of massive triplet ISFs and one tower of massless triplet ISFs}. First, we discuss a cubic vertex for CSF shown in \eqref{08082025-man03-01}, one tower of mass-$m_2$ triplet ISFs, and one tower of massless triplet ISFs. The general f-solution for the cubic vertex $\LL_\smp3$ is given by
\beq
\label{14082025-man03-02} && \hspace{-1cm} \LL_\smp3 =  L_1^{\SSc_1} V(L_2\,, \Qbf_{12}\,, Q_{23}\,, \Qbf_{31}; Q_{22},q_{33})\,,
\\
\label{14082025-man03-03} &&  L_1 =  p_3 \xi_1\,, \qquad L_2 =  p_1 \alpha_2 + \frac{m_1^2+m_2^2}{2m_2}\zeta_2 \,,\quad B_3 = p_2\alpha_3\,,
\nonumber\\
&& Q_{12}  =  \xi_1 \alpha_2 - \frac{\zeta_2}{m_2} L_1\,, \qquad  Q_{23}  =  \alpha_{23} + \frac{\zeta_2}{m_2} B_3 -  \frac{ 2 }{ m_1^2 - m_2^2 }  L_2 B_3\,,
\nonumber\\
&& Q_{31}  =  \alpha_3\xi_1 +  \frac{ 2 }{ m_1^2 - m_2^2 }  B_3 L_1\,,
\nonumber\\
&& \Qbf_{12}  = \frac{Q_{12}}{L_1}\,, \qquad \Qbf_{31}  = \frac{Q_{31}}{L_1}\,, \qquad Q_{22} = \alpha_{22} + \zeta_2\zeta_2\,, \qquad q_{33} = \alpha_{33}\,,
\eeq
where we use notation \eqref{17092025-man03-64} and introduce new vertex $V$ which depends on four variables shown explicitly in \eqref{14082025-man03-02} and defined in \eqref{14082025-man03-03}. Let us make the following remarks.

\noinbf{i)} The vertex $V$ is not fixed by equations \eqref{17092025-man03-10} and realized therefore as the freedom of our solution for the cubic vertex $\LL_\smp3$. This tells us that we have many cubic interaction vertices.

\noinbf{ii)} From  \eqref{14082025-man03-02}, \eqref{14082025-man03-03}, we note that
the dependence of the cubic vertex $\LL_\smp3$ on the momenta $p_a^\mu$ is governed by the variables $L_1$, and $L_2$, $\Qbf_{12}$, $Q_{23}$, $\Qbf_{31}$.

\noinbf{iii)} The cubic vertex $\LL_\smp3$ \eqref{14082025-man03-02} is non-polynomial in the momenta $p_a^\mu$ and therefore non-local. Namely, taking into account the expression for $\SSc$ in \eqref{16092025-01-man03-02} and the pre-factor $L_1^{ \SSc_1}$ \eqref{14082025-man03-02}, we note that the cubic vertex $\LL_\smp3$ \eqref{14082025-man03-02} is realized as $PL$-non-local cubic vertex (see \eqref{30032025-man01-35}).

\noinbf{iv}) The variables $L_2$, $\Qbf_{12}$, $Q_{23}$, $\Qbf_{31}$, $Q_{22}$, $q_{33}$ are realized as linear and quadratical forms in the oscillators $\alpha_2^\mu$, $\zeta_2$, $\alpha_3^\mu$. Therefore in order for the cubic vertex $\LL_\smp3$ \eqref{14082025-man03-02} to be sensible, this cubic vertex is required to be expandable in the just mentioned variables.

\noinbf{v)} To demonstrate explicitly the one-to-one correspondence of our covariant results with the light-cone gauge results in ref.\cite{Metsaev:2025qkr} we use notation \eqref{29092025-man03-01}, \eqref{29092025-man03-08} in appendix \ref{connect}. Using then the notation $X|_\lcgrm$ for the light-cone gauge projection of the variables in \eqref{14082025-man03-02}, we note the relations
\beq
\label{14082025-man03-10} && L_1|_\lcgrm = \ccc_1 L_1^u\,,  \qquad L_2|_\lcgrm =  L_2^\alpha\,,  \qquad \Qbf_{12}|_\lcgrm = \Qbf_{12}^{u\alpha}\,,
\nonumber\\
&&  Q_{23}|_\lcgrm =  Q_{23}^{\alpha\alpha}\,,  \qquad \Qbf_{31}|_\lcgrm = \Qbf_{31}^{\alpha u}\,,  \qquad Q_{22}|_\lcgrm = Q_{22}^{\alpha\alpha}\,,
\qquad q_{33}|_\lcgrm = q_{33}^{\alpha\alpha}\,,\qquad
\eeq
where expressions on r.h.s in \eqref{14082025-man03-10} enter the light-cone gauge cubic vertex in (6.8) in ref.\cite{Metsaev:2025qkr}.
The factor $\ccc_1$ in \eqref{14082025-man03-10} leads to the factor $E_1$ in \eqref{29092025-man03-15}.

\noinbf{One massive CSF, one massive triplet ISF and one massless triplet ISF}. For one CSFs shown in \eqref{14082025-man03-01}, one mass-$m_2$ and spin-$s_2$ triplet field, and one massless spin-$s_3$ triplet field, the cubic vertex $\LL_\smp3$ can be presented as in \eqref{14082025-man03-02}, \eqref{14082025-man03-03}, where the vertex $V$ is given by
{\small
\beq
\label{14082025-man03-04} && \hspace{-1.2cm} V = L_2^{k_2}  \Qbf_{12}^{n_3} Q_{23}^{n_1} \Qbf_{31}^{n_2} Q_{22}^{l_2} q_{33}^{l_3}\,,
\nonumber\\
&& \hspace{-0.3cm}   k_2 = s_2 - n_1 - n_3 - 2 l_2\,, \quad   n_1 +  n_2 + 2 l_3 = s_3\,,\quad k_2\,, n_1\,, n_2\,, n_3\,, l_2\,, l_3 \in \No_0\,.
\eeq
}
\!The integers $n_1$, $n_2$, $n_3$, $l_2$, $l_2$ subject to the conditions in \eqref{14082025-man03-04} describe a freedom of our solution for the cubic vertex $\LL_\smp3$.

\noinbf{One massive CSF, one massive ISF, and one massless ISF}.For fields shown in \eqref{14082025-man03-01}, cubic vertex $\LL_\smp3$ can be presented as in \eqref{14082025-man03-02}, \eqref{14082025-man03-03}, where the vertex $V$ is given by
{\small
\be
\label{14082025-man03-05} V = L_2^{k_2}  \Qbf_{12}^{n_3} Q_{23}^{n_1} \Qbf_{31}^{n_2}\big|_\Thsm\,,\qquad  \hspace{-0.3cm}   k_2 = s_2 - n_1 - n_3 \,, \quad  n_1 +  n_2 = s_3\,,\quad k_2\,, n_1\,, n_2\,, n_3 \in \No_0\,.
\ee
}
The integers $n_1$, $n_2$, $n_3$ subject to the conditions in \eqref{14082025-man03-05} describe a freedom of the solution for the cubic vertex $\LL_\smp3$. The notation $|_\Thsm$ indicates that, in \eqref{14082025-man03-05}, we make the replacement $\alpha_2^\mu\rightarrow \alpha_{\Thsm\,2}^\mu$, $\zeta_2\rightarrow \zeta_{\Thsm\,2}$, $\alpha_3^\mu\rightarrow \alpha_{\Thsm\,3}^\mu$, where the new oscillators $\alpha_{\Thsm\,2}^\mu$, $\zeta_{\Thsm,2}$ and $\alpha_{\Thsm\,3}^\mu$ are defined as in \eqref{20082025-man03-05} and  \eqref{20082025-man03-06} in appendix \ref{app-notation}. In the expression for the action \eqref{17092025-man03-20}, such replacement can be ignored (see our remark at the end of section \ref{sec-03}).

\subsection{ \large One massive CSF and two massless ISFs}

We consider the cubic vertex for one massive CSFs and two massless ISFs:
\beq
\label{15082025-man03-01} && (m_1,\SSc_1)_\CSFsm\hbox{-}(0,s_2)_\ISFsm \hbox{-}(0,s_3)_\ISFsm \,,\qquad m_1^2 < 0 \,,
\nonumber\\
&&  \hbox{\small one massive CSF and two massless ISFs.}\quad
\eeq
\noinbf{One massive CSF and two towers of massless triplet ISFs}. First, we discuss a cubic vertex for CSF shown in \eqref{15082025-man03-01} and two towers of massless triplet ISFs. The general f-solution for the cubic vertex $\LL_\smp3$ is given by
\beq
\label{15082025-man03-02} && \hspace{-1cm} \LL_\smp3 =  L_1^{\SSc_1} V(\Qbf_{12}\,, Q_{23}\,, \Qbf_{31};q_{22}, q_{33})\,,
\\
\label{15082025-man03-03} &&  L_1 =  p_3 \xi_1\,,\quad B_2 = p_1 \alpha_2\,, \quad B_3 = p_2\alpha_3\,,
\nonumber\\
&& Q_{12}  =  \xi_1\alpha_2  +  \frac{2}{m_1^2} L_1 B_2 \,,
\nonumber\\
&& Q_{23} =  \alpha_{23}  -  \frac{ 2 }{ m_1^2} B_2 B_3\,,  \quad Q_{31} =  \alpha_3\xi_1    +\frac{ 2 }{ m_1^2} B_3 L_1\,,
\nonumber\\
&& \Qbf_{12}  = \frac{Q_{12}}{L_1}\,, \quad \Qbf_{31}  = \frac{Q_{31}}{L_1}\,, \quad q_{22}=\alpha_{22},\quad q_{33}=\alpha_{33}\,,
\eeq
where we use notation \eqref{17092025-man03-64} and introduce new vertex $V$ which depends on four variables shown explicitly in \eqref{15082025-man03-02} and defined in \eqref{15082025-man03-03}. Let us make the following remarks.

\noinbf{i)} The vertex $V$ is not fixed by equations \eqref{17092025-man03-10} and realized therefore as the freedom of our solution for the cubic vertex $\LL_\smp3$. This tells us that we have many cubic interaction vertices.

\noinbf{ii)} From  \eqref{15082025-man03-02}, \eqref{15082025-man03-03}, we note that
the dependence of the cubic vertex $\LL_\smp3$ on the momenta $p_a^\mu$ is governed by the variables $L_1$ and $\Qbf_{12}$, $Q_{23}$, $\Qbf_{31}$.

\noinbf{iii)} The cubic vertex $\LL_\smp3$ \eqref{15082025-man03-02} is non-polynomial in the momenta $p_a^\mu$ and therefore non-local. Namely, taking into account the expression for $\SSc$ in \eqref{16092025-01-man03-02} and the pre-factor $L_1^{ \SSc_1}$ \eqref{15082025-man03-02}, we note that the cubic vertex $\LL_\smp3$ \eqref{15082025-man03-02} is realized as $PL$-non-local cubic vertex (see \eqref{30032025-man01-35}).

\noinbf{iv}) The variables $\Qbf_{12}$, $Q_{23}$, $\Qbf_{31}$, $q_{22}$, $q_{33}$ are realized as linear and quadratical forms in the oscillators $\alpha_2^\mu$, $\alpha_3^\mu$. Therefore in order for the cubic vertex $\LL_\smp3$ \eqref{15082025-man03-02} to be sensible, this cubic vertex is required to be expandable in the just mentioned variables.

\noinbf{v)} To demonstrate explicitly the one-to-one correspondence of our covariant results with the light-cone gauge results in ref.\cite{Metsaev:2025qkr} we use notation \eqref{29092025-man03-01}, \eqref{29092025-man03-08} in appendix \ref{connect}. Using then the notation $X|_\lcgrm$ for the light-cone gauge projection of the variables in \eqref{15082025-man03-02}, we note the relations
\beq
\label{15082025-man03-10} && L_1|_\lcgrm = \ccc_1 L_1^u\,,  \qquad \Qbf_{12}|_\lcgrm = \Qbf_{12}^{u\alpha}\,, \qquad Q_{23}|_\lcgrm =  Q_{23}^{\alpha\alpha}\,,
\nonumber\\
&&  \Qbf_{31}|_\lcgrm = \Qbf_{31}^{\alpha u}\,,  \qquad q_{22}|_\lcgrm = q_{22}^{\alpha\alpha}\,,
\qquad q_{33}|_\lcgrm = q_{33}^{\alpha\alpha}\,,\qquad
\eeq
where expressions on r.h.s in \eqref{15082025-man03-10} enter the light-cone gauge cubic vertex in (6.14) in ref.\cite{Metsaev:2025qkr}.
The factor $\ccc_1$ in \eqref{15082025-man03-10} leads to the factor $E_1$ in \eqref{29092025-man03-15}.

\noinbf{One massive CSF and two massless triplet ISFs}. For one CSFs shown in \eqref{15082025-man03-01}, one massless spin-$s_2$ triplet field, and one massless spin-$s_3$ triplet field, the cubic vertex $\LL_\smp3$ can be presented as in \eqref{15082025-man03-02}, \eqref{15082025-man03-03}, where the vertex $V$ is given by
{\small
\be
\label{15082025-man03-04} V = \Qbf_{12}^{n_3} Q_{23}^{n_1} \Qbf_{31}^{n_2} q_{22}^{l_2} q_{33}^{l_3}\,,\quad     s_2 = n_1 + n_3 + 2 l_2\,, \quad  s_3 =  n_1 + n_2 + 2 l_3\,,\quad  n_1\,, n_2\,, n_3\,, l_2\,, l_3 \in \No_0\,.
\ee
}
\!The integers $n_1$, $n_2$, $n_3$, $l_2$, $l_3$ subject to the conditions in \eqref{15082025-man03-04} describe a freedom of our solution for the cubic vertex $\LL_\smp3$.

\noinbf{One massive CSF and two massless ISFs}.
For fields shown in \eqref{15082025-man03-01}, cubic vertex $\LL_\smp3$ can be presented as in \eqref{15082025-man03-02}, \eqref{15082025-man03-03}, where the vertex $V$ is given by
{\small
\be \label{15082025-man03-05}
V = \Qbf_{12}^{n_3} Q_{23}^{n_1} \Qbf_{31}^{n_2}\big|_\Thsm\,,\quad s_2 =  n_1 + n_3 \,, \quad s_3 =  n_1 + n_2\,,\quad  n_1\,, n_2\,, n_3 \in \No_0\,.
\ee
}
\!The integers $n_1$, $n_2$, $n_3$ subject to the conditions in \eqref{15082025-man03-05} describes a freedom of the solution for the cubic vertex $\LL_\smp3$. The notation $|_\Thsm$ indicates that, in \eqref{15082025-man03-05}, we make the replacement $\alpha_a^\mu\rightarrow \alpha_{\Thsm\,a}^\mu$, $a=1,2$, where the new oscillators $\alpha_\Thsm^\mu$ are given in \eqref{20082025-man03-06} in appendix \ref{app-notation}. In the expression for the action \eqref{17092025-man03-20}, such replacement can be ignored (see our remark at the end of section \ref{sec-03}).

\subsection{  \large One massless CSF and two massive ISFs (non-equal masses)}

We consider the cubic vertex for one massless CSF  and two massive ISFs:
\beq
\label{16082025-man03-01} && (0,\kappa_1)_\CSFsm\hbox{-}(m_2,s_2)_\ISFsm \hbox{-}(m_3,s_3)_\ISFsm \,, \qquad m_2^2 > 0\qquad m_3^2>0\,,\qquad m_2^2 \ne m_3^2\,,
\nonumber\\
&&  \hbox{\small one massless CSF and two massive ISFs with non-equal masses.}\quad
\eeq
\noinbf{One massless CSF and two towers of massive triplet ISFs}. First, we discuss a cubic vertex for CSF shown in \eqref{16082025-man03-01}, one tower of mass-$m_2$ triplet ISFs, and one tower of mass-$m_3$ triplet ISFs, $m_2^2\ne m_3^2$. The general f-solution for the cubic vertex $\LL_\smp3$ is given by
\beq
\label{16082025-man03-02} && \hspace{-1cm} \LL_\smp3 =   e^W V(L_2,L_3,Q_{12}\,, Q_{23}\,, Q_{31};\, Q_{22}\,, Q_{33})\,,
\\
\label{16082025-man03-03} && B_1 = p_3\xi_1\,,\quad  L_2 =  p_1 \alpha_2  + \frac{m_2^2-m_3^2}{2m_2}\zeta_2\,, \quad  L_3 =  p_2 \alpha_3 + \frac{m_2^2+m_3^2}{2m_3}\zeta_3\,,
\nonumber\\
&& W = -\frac{2\irm }{m_2^2 - m_3^2} B_1 \,,
\nonumber\\
&& Q_{12}  =  \xi_1 \alpha_2 - \frac{\zeta_2}{m_2}B_1 +  \frac{2}{m_2^2-m_3^2} B_1 L_2 \,,
\nonumber\\
&& Q_{23}  =  \alpha_{23} + \frac{\zeta_2}{m_2} L_3 - \frac{\zeta_3}{m_3} L_2 - \frac{m_2^2 + m_3^2}{2m_2m_3}\zeta_2\zeta_3\,,
\nonumber\\
&& Q_{31} =  \alpha_3\xi_1 + \frac{\zeta_3}{m_3} B_1 -  \frac{ 2 }{ m_2^2 - m_3^2} L_3 B_1\,, \qquad  Q_{aa} = \alpha_{aa} + \zeta_a\zeta_a\,, \quad a=2,3\,,\qquad
\eeq
where we use notation \eqref{17092025-man03-64} and introduce new vertex $V$ which depends on four variables shown explicitly in \eqref{16082025-man03-02} and defined in \eqref{16082025-man03-03}. Let us make the following remarks.

\noinbf{i)} The vertex $V$ is not fixed by equations \eqref{17092025-man03-10} and realized therefore as the freedom of our solution for the cubic vertex $\LL_\smp3$. This tells us that we have many cubic interaction vertices.

\noinbf{ii)} From  \eqref{16082025-man03-02}, \eqref{16082025-man03-03}, we note that
the dependence of the cubic vertex $\LL_\smp3$ on the momenta $p_a^\mu$ is governed by the variables $W$, $L_2$, $L_3$, $Q_{12}$, $Q_{23}$, $Q_{31}$.

\noinbf{iii)} The cubic vertex $\LL_\smp3$ \eqref{16082025-man03-02} is non-polynomial in the momenta $p_a^\mu$ and therefore non-local. Namely, taking into account the pre-factor $e^W$ \eqref{16082025-man03-02}, we note that the cubic vertex $\LL_\smp3$ \eqref{16082025-man03-02} is realized as $E$-non-local cubic vertex (see \eqref{30032025-man01-35}).

\noinbf{iv}) The variables $L_2$, $L_3$ $Q_{12}$, $Q_{23}$, $Q_{31}$, $Q_{22}$, $Q_{33}$ are realized as linear and quadratical forms in the oscillators $\alpha_2^\mu$, $\zeta_2$, $\alpha_3^\mu$, $\zeta_3$. Therefore in order for the cubic vertex $\LL_\smp3$ \eqref{16082025-man03-02} to be sensible, this cubic vertex is required to be expandable in the just mentioned variables.

\noinbf{v)} To demonstrate explicitly the one-to-one correspondence of our covariant results with the light-cone gauge results in ref.\cite{Metsaev:2025qkr} we use notation \eqref{29092025-man03-01}, \eqref{29092025-man03-08} in appendix \ref{connect}. Using then the notation $X|_\lcgrm$ for the light-cone gauge projection of the variables in \eqref{16082025-man03-02}, we note the relations
\beq
\label{16082025-man03-10} && W|_\lcgrm = W + \frac{\irm \xi_1^+}{\beta_1}\,,  \qquad L_2|_\lcgrm =  L_2^\alpha\,, \qquad  L_3|_\lcgrm = L_3^\alpha\,,\qquad Q_{12}|_\lcgrm = c_1 Q_{12}^{u\alpha}\,,
\nonumber\\
&& Q_{23}|_\lcgrm =  Q_{23}^{\alpha\alpha}\,,  \qquad Q_{31}|_\lcgrm = c_1Q_{31}^{\alpha u}\,,  \qquad Q_{22}|_\lcgrm = Q_{22}^{\alpha\alpha}\,,
\qquad Q_{33}|_\lcgrm = Q_{33}^{\alpha\alpha}\,,\qquad
\eeq
where expressions on r.h.s in \eqref{16082025-man03-10} enter the light-cone gauge cubic vertex in (6.19) in ref.\cite{Metsaev:2025qkr}.
The $\xi_1^+$-term in \eqref{16082025-man03-10} leads to the factor $E_1$ in \eqref{29092025-man03-15}.

\noinbf{One massless CSF and two massive triplet ISFs}. For one CSFs shown in \eqref{16082025-man03-01}, one mass-$m_2$ and spin-$s_2$ triplet field, and one mass-$m_3$ and spin-$s_3$ triplet field, $m_2^2\ne m_3^2$, the cubic vertex $\LL_\smp3$ can be presented as in \eqref{16082025-man03-02}, \eqref{16082025-man03-03}, where the vertex $V$ is given by
{\small
\beq
\label{16082025-man03-04} && \hspace{-1.2cm} V = L_2^{k_2} L_3^{k_3} Q_{12}^{n_3} Q_{23}^{n_1} Q_{31}^{n_2} Q_{22}^{l_2} Q_{33}^{l_3}\,,
\nonumber\\
&& \hspace{-0.3cm}   k_2 = s_2 - n_1 - n_3 - 2 l_2\,, \quad k_3  = s_3 - n_1 - n_2 - 2 l_3\,,\quad k_2\,, k_3\,, n_1\,, n_2\,, n_3\,, l_2\,, l_3 \in \No_0\,.\qquad
\eeq
}
\!The five integers $n_1$, $n_2$, $n_3$, $l_2$, $l_3$ subject to the conditions in \eqref{16082025-man03-04} describe a freedom of our solution for the cubic vertex $\LL_\smp3$.

\noinbf{One massless CSF and two massive ISF}. For fields shown in \eqref{16082025-man03-01}, cubic vertex $\LL_\smp3$ can be presented as in \eqref{16082025-man03-02}, \eqref{16082025-man03-03}, where the vertex $V$ is given by
{\small
\be
\label{16082025-man03-05}   V = L_2^{k_2} L_3^{k_3} Q_{12}^{n_3} Q_{23}^{n_1} Q_{31}^{n_2}\big|_\Thsm\,, \qquad k_2 = s_2 - n_1 - n_3 \,, \quad k_3  = s_3 - n_1 - n_2\,,\quad k_2\,, k_3\,, n_1\,, n_2\,, n_3 \in \No_0\,.
\ee
}
The three integers $n_1$, $n_2$, $n_3$ subject to the conditions in \eqref{16082025-man03-05} describe a freedom of the solution for the cubic vertex $\LL_\smp3$. The notation $|_\Thsm$ indicates that, in \eqref{16082025-man03-05}, we make the replacement $\alpha_a^\mu\rightarrow \alpha_{\Thsm\,a}^\mu$, $\zeta_a\rightarrow \zeta_{\Thsm\,a}$, $a=1,2$, where the new oscillators $\alpha_\Thsm^\mu$, $\zeta_\Thsm$  are given in \eqref{20082025-man03-05} in appendix \ref{app-notation}. In the expression for the action \eqref{17092025-man03-20}, such replacement can be ignored (see our remark at the end of section \ref{sec-03}).

\subsection{  One massless CSF and two massive ISFs (equal masses)} \label{subsec-6-5}

We consider the cubic vertex for one massive CSFs  and two massive ISFs:
\beq
\label{17082025-man03-01}  && (0,\kappa_1)_\CSFsm\hbox{-}(m_2,s_2)_\ISFsm \hbox{-}(m_3,s_3)_\ISFsm \,,\qquad m_2^2=m^2\,, \qquad m_3^2=m^2\,, \qquad m^2 > 0\,,
\nonumber\\
&&  \hbox{\small one massless CSF and two massive ISFs with equal masses.}\quad
\eeq
We make the following comments.

\noinbf{\ibf)} We verified that equations for cubic interaction vertex \eqref{17092025-man03-10} do not admit f-solutions. In other words, Lorentz covariant approach we use here does not admit f-solutions to cubic vertices for fields in \eqref{17082025-man03-01}. This is in agreement with the conclusion in ref.\cite{Metsaev:2025qkr} that the light-cone gauge vector superspace formulation also does not admit f-solutions for  fields in \eqref{17082025-man03-01}.

\noinbf{\iibf)} In ref.\cite{Metsaev:2017cuz}, using the oscillator light-cone gauge approach, we find the cubic vertices for fields shown in \eqref{17082025-man03-01}. In ref.\cite{Metsaev:2025qkr}, using transformation from light-cone gauge oscillator formulation to light-cone gauge vector superspace formulation,  we shown that such cubic vertices are realized as distributional cubic vertices in light-cone gauge vector superspace.  At the present time, we do know how to find Lorentz covariant d-solution corresponding to the light-cone gauge d-solution in section 6.5 in ref.\cite{Metsaev:2025qkr} Note also we have no proof that d-solution for Lorentz covariant cubic vertex for fields \eqref{17082025-man03-01} does not exist.

\subsection{  \large One massless CSF, one massive ISF and one massless ISF}

We consider the cubic vertex for one massive CSFs  and two massive ISFs:
\beq
\label{18082025-man03-01} && (0,\kappa_1)_\CSFsm\hbox{-}(m_2,s_2)_\ISFsm \hbox{-}(0,s_3)_\ISFsm \,,\qquad m_2^2 > 0\,,
\nonumber\\
&&  \hbox{\small one massless CSF, one massive ISF, and one massless ISF.}\quad
\eeq
\noinbf{One massless CSF, one tower of massive triplet ISFs and one tower of massless triplet ISFs}. First, we discuss a cubic vertex for CSF shown in \eqref{18082025-man03-01}, one tower of mass-$m_2$ triplet ISFs, and one tower of massless triplet ISFs. The general f-solution for the cubic vertex $\LL_\smp3$ is given by
\beq
\label{18082025-man03-02} && \hspace{-1cm} \LL_\smp3 =   e^W  V(L_2\,,Q_{12}\,, Q_{23}\,, Q_{31};\, Q_{22}\,, q_{33})\,,
\\
\label{18082025-man03-03} && B_1=p_3\xi_1\,, \quad L_2 =  p_1 \alpha_2 + \half m_2 \zeta_2 \,,\quad B_3 = p_2\alpha_3\,,\quad W  = - \frac{2\irm }{m_2^2} B_1\,,
\nonumber\\
&& Q_{12}  =  \xi_1 \alpha_2 - \frac{\zeta_2}{m_2} B_1  + \frac{2}{m_2^2} B_1 L_2 \,, \hspace{1cm}  Q_{23}  =  \alpha_{23} + \frac{\zeta_2}{m_2} B_3 +  \frac{2}{m_2^2} L_2 B_3\,,
\nonumber\\
&& Q_{31}  =  \alpha_3\xi_1 - \frac{ 2 }{ m_2^2 }  B_3 B_1\,,
\hspace{2.6cm} Q_{22} = \alpha_{22} + \zeta_2\zeta_2\,,\quad q_{33}=\alpha_{33}\,,
\eeq
where we use notation \eqref{17092025-man03-64} and introduce new vertex $V$ which depends on six variables shown explicitly in \eqref{18082025-man03-02} and defined in \eqref{18082025-man03-03}. Let us make the following remarks.

\noinbf{i)} The vertex $V$ is not fixed by equations \eqref{17092025-man03-10} and realized therefore as the freedom of our solution for the cubic vertex $\LL_\smp3$. This tells us that we have many cubic interaction vertices.

\noinbf{ii)} From  \eqref{18082025-man03-02}, \eqref{18082025-man03-03}, we note that
the dependence of the cubic vertex $\LL_\smp3$ on the momenta $p_a^\mu$ is governed by the variables $W$,  $L_2$, $Q_{12}$, $Q_{23}$, $Q_{31}$.

\noinbf{iii)} The cubic vertex $\LL_\smp3$ \eqref{18082025-man03-02} is non-polynomial in the momenta $p_a^\mu$ and therefore non-local. Namely, taking into account the expression for the pre-factor $e^W$ \eqref{18082025-man03-02}, we note that the cubic vertex $\LL_\smp3$ \eqref{18082025-man03-02} is realized as $E$-non-local cubic vertex (see \eqref{30032025-man01-35}).

\noinbf{iv}) The variables $L_2$, $Q_{12}$, $Q_{23}$, $Q_{31}$, $Q_{22}$, $q_{33}$, are realized as linear and quadratical forms in the oscillators $\alpha_2^\mu$, $\zeta_2$, $\alpha_3^\mu$. Therefore in order for the cubic vertex $\LL_\smp3$ \eqref{18082025-man03-02} to be sensible, this cubic vertex is required to be expandable in the just mentioned variables.

\noinbf{v)} To demonstrate explicitly the one-to-one correspondence of our covariant results with the light-cone gauge results in ref.\cite{Metsaev:2025qkr} we use notation \eqref{29092025-man03-01}, \eqref{29092025-man03-08} in appendix \ref{connect}. Using then the notation $X|_\lcgrm$ for the light-cone gauge projection of the variables in \eqref{18082025-man03-02}, we note the relations
\beq
\label{18082025-man03-10} && W|_\lcgrm = W + \frac{\irm \xi_1^+}{\beta_1}\,,  \qquad L_2|_\lcgrm =  L_2^\alpha\,, \qquad   Q_{12}|_\lcgrm = c_1 Q_{12}^{u\alpha}\,,
\nonumber\\
&& Q_{23}|_\lcgrm =  Q_{23}^{\alpha\alpha}\,,  \qquad Q_{31}|_\lcgrm = c_1Q_{31}^{\alpha u}\,,  \qquad Q_{22}|_\lcgrm = Q_{22}^{\alpha\alpha}\,,
\qquad q_{33}|_\lcgrm = q_{33}^{\alpha\alpha}\,,\qquad
\eeq
where expressions on r.h.s in \eqref{18082025-man03-10} enter the light-cone gauge cubic vertex in (6.30) in ref.\cite{Metsaev:2025qkr}.
The $\xi_1^+$-term in \eqref{18082025-man03-10} leads to the factor $E_1$ in \eqref{29092025-man03-15}.

\noinbf{One massless CSF, one massive triplet ISF and one massless triplet ISF}.  For one CSFs shown in \eqref{18082025-man03-01}, one mass-$m_2$ and spin-$s_2$ triplet field, and one massless spin-$s_3$ triplet field, the cubic vertex $\LL_\smp3$ can be presented as in \eqref{18082025-man03-02}, \eqref{18082025-man03-03}, where the vertex $V$ is given by
{\small
\beq
\label{18082025-man03-04} && \hspace{-1.2cm} V = L_2^{k_2} Q_{12}^{n_3} Q_{23}^{n_1} Q_{31}^{n_2} Q_{22}^{l_2} q_{33}^{l_3}\,,
\nonumber\\
&& \hspace{-0.3cm}   k_2 = s_2 - n_1 - n_3 - 2 l_2\,, \quad   n_1 +  n_2 + 2 l_3 = s_3\,,\quad k_2\,, n_1\,, n_2\,, n_3\,, l_2\,, l_3 \in \No_0\,.
\eeq
}
\!The integers $n_1$, $n_2$, $n_3$, $l_2$, $l_3$ subject to the conditions in \eqref{18082025-man03-04} describe a freedom of our solution for the cubic vertex $\LL_\smp3$.

\noinbf{One massless CSF, one massive ISF, and one massless ISF}. For fields shown in \eqref{18082025-man03-01}, cubic vertex $\LL_\smp3$ can be presented as in \eqref{18082025-man03-02}, \eqref{18082025-man03-03}, where the vertex $V$ is given by
{\small
\be
\label{18082025-man03-05}   V = L_2^{k_2}  Q_{12}^{n_3} Q_{23}^{n_1} Q_{31}^{n_2}\big|_\Thsm\,, \qquad k_2 = s_2 - n_1 - n_3 \,, \quad  n_1 +  n_2 = s_3\,,\quad k_2\,, n_1\,, n_2\,, n_3 \in \No_0\,.
\ee
}
\!The integers $n_1$, $n_2$, $n_3$ subject to the conditions in \eqref{18082025-man03-05} describe a freedom of the solution for the cubic vertex $\LL_\smp3$. The notation $|_\Thsm$ indicates that, in \eqref{18082025-man03-05}, we make the replacement $\alpha_2^\mu\rightarrow \alpha_{\Thsm\,2}^\mu$, $\zeta_2\rightarrow \zeta_{\Thsm\,2}$ and $\alpha_3^\mu\rightarrow \alpha_{\Thsm\,3}^\mu$, where the new oscillators are given in \eqref{20082025-man03-05} and \eqref{20082025-man03-06} in appendix \ref{app-notation}. In the expression for the action \eqref{17092025-man03-20}, such replacement can be ignored (see our remark at the end of section \ref{sec-03}).

\subsection{  One massless CSF and two massless ISFs}

We consider the cubic vertex for one massive CSFs  and two massless ISFs:
\beq
\label{19082025-man03-01} && (0,\kappa_1)_\CSFsm\hbox{-}(0,s_2)_\ISFsm \hbox{-}(0,s_3)_\ISFsm \,,
\nonumber\\
&&  \hbox{\small one massless CSF and two massless ISF.}\quad
\eeq
We make the following comments.

\noinbf{\ibf)} We verified that equations for cubic interaction vertex \eqref{17092025-man03-10} do not admit f-solutions. In other words, Lorentz covariant approach we use here does not admit f-solutions to cubic vertices for fields in \eqref{19082025-man03-01}. This is in agreement with the conclusion in  ref.\cite{Metsaev:2025qkr} that the light-cone gauge vector superspace approach  also does not admit f-solutions to cubic vertex for fields in \eqref{19082025-man03-01}.

\noinbf{\iibf)} As in the case of light-cone gauge vector superspace formulation, we have no proof that equations for Lorentz covariant cubic vertex for fields \eqref{17082025-man03-01} do not admit d-solutions.

\section{Conclusions}\label{concl}

In this paper, we used on-shell Lorentz covariant formulation of relativistic dynamics for a study of cubic vertices of CSFs and ISFs which involve at least one CSF.
We presented solution to all parity-even cubic vertices which are realized as functions on the Lorentz covariant vector superspace and build the corresponding manifestly Lorentz invariant on-shell cubic action. We shown that such action suffers from divergencies and, for this reason, the manifestly Lorentz invariant on-shell cubic action is referred to as formal action in our study. We find the modification of the formal action which removes the divergencies and maintains the Lorentz symmetries. We demonstrated that the modified on-shell action considered in the light-cone gauge coincides with the off-shell light-cone gauge action evaluated on the solution of free light-cone gauge equation of motions. This provides an additional check for our solution to the on-shell cubic action.

Our results provide the on-shell Lorentz covariantization of our light-cone gauge cubic vertices for CSFs and ISFs obtained in ref.\cite{Metsaev:2025qkr}. For all f-solutions and some particular d-solutions, we demonstrated explicitly the one-to-one correspondence of the light-cone gauge cubic vertices and the corresponding on-shell Lorentz covariant cubic vertices. As compared to the light-cone gauge cubic vertices the on-shell Lorentz covariant cubic vertices turn out to be simpler. We think  therefore that our Lorentz covariant vertices can be considered as a promising and good starting point for building off-shell Lagrangian Lorentz covariant formulation of interacting CSFs. We believe that our results might be helpful for the following generalizations and applications.

\noinbf{\ibf)} BRST approach is a powerful tool for investigating off-shell formulation of relativistic interacting fields. In the framework of BRST approach, a free CSFs propagating in flat space have been studied in ref.\cite{Bengtsson:2013vra,Metsaev:2018lth,Buchbinder:2018yoo, Buchbinder:2020nxn,Burdik:2020ror}. Needless to say that our manifestly Lorentz covariant cubic vertices might be helpful for finding their BRST cousins. For BRST studies of interacting ISFs, see, e.g., refs.\cite{Bengtsson:1987jt,Bekaert:2005jf,Fotopoulos:2010ay,Metsaev:2012uy, Grigoriev:2020lzu,Buchbinder:2021xbk}. The discussion of a free mixed-symmetry CSF by using BRST method may be found in refs.\cite{Alkalaev:2017hvj}.%
\footnote{For light-cone gauge studies of cubic vertices for ISFs, see refs.\cite{Bengtsson:1986kh,Metsaev:2007rn,Metsaev:2022yvb}, while metric-like formulations of cubic vertices were considered in refs.\cite{Manvelyan:2010jr,Sagnotti:2010at}.}

\noinbf{\iibf)} In recent time, CSF propagating in AdS space has actively been investigated in literature. For CSF in AdS, the metric-like formulation was obtained ref.\cite{Metsaev:2016lhs,Metsaev:2017ytk}, while frame-like formulation was developed in ref.\cite{Zinoviev:2017rnj}. Light-cone gauge oscillator formulation of CSF in AdS was investigated in refs.\cite{Metsaev:2017myp,Metsaev:2019opn,Metsaev:2021zdg}, while BRST studies of  CSF in AdS may be found in refs.\cite{Buchbinder:2024jpt,Buchbinder:2024vli}. Mixed-symmetry CSF in AdS was studied
in refs.\cite{Khabarov:2017lth,Metsaev:2017myp,Metsaev:2021zdg,Golubtsova:2025eei}, while
unfolded formulation of CSF in AdS was developed in ref.\cite{Khabarov:2020glf} (see also ref.\cite{Ponomarev:2010st}). In ref.\cite{Metsaev:2025nbm}, we generalized the light-cone gauge vector superspace formulation to the case of CSF in AdS. We believe that results in ref.\cite{Metsaev:2025nbm} and the ones obtained in this paper will be useful for study of interacting CSF in AdS.

\noinbf{\iiibf)} We expect that a worldline approach to a continuous spin will provide new interesting possibilities for the study of CSF. Continuous-spin (super)particle has been studied in ref.\cite{Buchbinder:2018soq} (see also ref.\cite{Basile:2023vyg}). For the recent interesting development of a continuous-spin superparticle in flat/curved superspace, see Refs.\cite{Buchbinder:2025ztn,Buchbinder:2025gsr}. It will be interesting to understand the way in which our interaction vertices are realized in the framework of worldline approach.%
\footnote{In the framework of universal model of spinning particle, a continuous spin was studied in ref.\cite{Lyakhovich:1996we}.  We cordially thank Prof. S.L. Lyakhovich for informing us about ref.\cite{Lyakhovich:1996we}.}

\noinbf{\ivbf)} Supersymmetric theories of interacting CSF is an interesting direction to go.
As is well known, in general, supersymmetry imposes additional restrictions on vertices and eliminates many a priori allowed vertices from the game. We expect that a supersymmetrization of our Lorentz covariant cubic vertices decreases a number of allowed vertices and hence simplifies a study of interacting CSF. Lagrangian formulation of free supersymmetric CSFs has been studied in refs.\cite{Najafizadeh:2019mun,Buchbinder:2019kuh}.%
\footnote{For $\NN=1$, all supersymmetric light-cone gauge cubic vertices for massless ISFs in $\Ro^{3,1}$ were obtained in ref.\cite{Metsaev:2019dqt}. Light-cone cubic vertices for $\NN=4\No$-extended massless scalar supermultiplet were studied in ref.\cite{Bengtsson:1983pg}, while light-cone cubic vertices for {\it arbitrary spin} (integer and half-integer) $\NN=4\No$-extended massless supermultiplets were investigated ref.\cite{Metsaev:2019aig}. For recent interesting study of $N=1$ SUSY in higher dimensions, see ref.\cite{Buchbinder:2021qrg}.
}

\appendix
\section{Notation and conventions  }\label{app-notation}

\noinbf{Notation}. The Lorentz algebra $so(d-1,1)$ vector indices take values $\mu,\nu=0,1,\ldots ,d-1$. We refer creation operators $\alpha^\mu$, $\zeta$ and the respective annihilation operators $\alphab^\mu$, $\zetab$ to as oscillators. We use the following conventions for the commutators, the vacuum $|0\rangle $, and hermitian conjugation rules:
\be
\label{20082025-man03-01} [ \alphab^\mu,\alpha^\nu] = \eta^{\mu\nu}, \quad [\zetab,\zeta]=1,  \quad \alphab^\mu |0\rangle = 0\,,  \quad \zetab |0\rangle = 0\,,\qquad \alpha^{\mu \dagger} = \alphab^\mu\,, \hspace{1.1cm} \zeta^\dagger = \zetab\,,
\ee
and the following shortcuts for scalar products:
\be \label{20082025-man03-02}
\alpha^2 := \alpha^\mu\alpha^\mu\,,\qquad \alphab^2 := \alphab^\mu \alphab^\mu\,,\qquad N_\alpha  := \alpha^\mu \alphab^\mu\,, \qquad N_\zeta  := \zeta \zetab\,,
\ee
where, in \eqref{20082025-man03-02} and throughout this paper, in the scalar products of Lorentz vectors, we skip mostly positive flat metric $\eta_{\mu\nu}$ and use the shortcut for a scalar product  $X^\mu Y^\mu:= \eta_{\mu\nu} X^\mu Y^\nu$.

Two sets of oscillators $\alpha_\Thsm^\mu$, $\zeta_\Thsm$ and $\alpha_\Thsm^\mu$ corresponding to the respective massive and massless fields are defined by the following relations:
{\small
\beq
\label{20082025-man03-05} &&  \alpha_\Thsm^\mu := \alpha^\mu - (\alpha^2+\zeta^2)\frac{1}{2N_\alpha + 2 N_\zeta + d+1}\alphab^\mu\,, \hspace{0.8cm}
\nonumber\\
&&  \zeta_\Thsm := \zeta - (\alpha^2+\zeta^2)\frac{1}{2N_\alpha + 2 N_\zeta + d+1}\zetab\,, \hspace{1.3cm} \hbox{for massive ISF},
\\
\label{20082025-man03-06} &&  \alpha_\Thsm^\mu := \alpha^\mu - \alpha^2\frac{1}{2N_\alpha + d}\alphab^\mu\,, \hspace{3.7cm} \hbox{for massless ISF}.\qquad
\eeq
}
The main property of the two set of oscillators is summarized as follows. Let $P_\msv(\alpha_\Thsm^\mu,\zeta_\Thsm)$ be a polynomial of the oscillators $\alpha_\Thsm^\mu$, $\zeta_\Thsm$, while $P_\msl(\alpha_\Thsm^\mu)$ be a polynomial of the oscillators  $\alpha_\Thsm^\mu$. We note then the following  relations:
{\small
\be \label{20082025-man03-10}
(\alphab^2 + \zetab^2) P_\msv(\alpha_\Thsm^\mu,\zeta_\Thsm)|0\rangle =  0\,, \quad \hbox{and}  \quad \alphab^2 P_\msl(\alpha_\Thsm^\mu)|0\rangle = 0\,.
\ee
}
An inner product for any quantities $A(\alpha)$ and $B(\alpha)$ that depend on the oscillators are defined as
{\small
\be \label{20082025-man03-15}
A(\alpha) \cdot B(\alpha) : = \langle A(\alpha)| B(\alpha)\rangle\,, \qquad \langle A(\alpha)| B(\alpha)\rangle : = \langle 0 | (A(\alpha))^\dagger B(\alpha)|0\rangle\,.\qquad
\ee
}

\noinbf{Light-cone frame notation}. In the light-cone frame, a Lorentz vector $X^\mu$ is decomposed as $X^\mu = X^+,X^-,X^I$, where vector indices of the $so(d-2)$ algebra take values $I,J=1,\ldots,d-2$, while the $X^+$, $X^-$ are defined as $X^\pm := (X^{d-1}  \pm X^0)/\sqrt{2}$. In such frame, the non vanishing elements of the flat metric are given by $\eta_{+-} = \eta_{-+}=1$, $\eta_{IJ} = \delta_{IJ}$ and hence the scalar product of two Lorentz vectors is decomposed as $X^\mu Y^\mu = X^+ Y^- + X^- Y^+ + X^I Y^I$, where $X^\mu Y^\mu:= \eta_{\mu\nu} X^\mu Y^\nu$.

\section{Light-cone frame for CSF and light-cone gauge for ISF}\label{div}

Here, for the reader convenience, we discuss a representation of relativistic CSF and ISF in terms of the respective constraints-free light-cone frame CSF and light-cone gauge ISF. The representation for the relativistic massive/massless CSF given below was obtained by using WBB-constraints \eqref{16092025-01-man03-01}, \eqref{16092025-01-man03-02} in appendices B,C in ref.\cite{Metsaev:2025qkr}.

\noinbf{Relativistic and light-cone frame massive/massless CSF}. Relativistic massive/massless CSF denoted as $\phi=\phi(p^\mu,\xi^\mu)$ depend on the momenta $p^\mu$ and the vector $\xi^\mu$, while a light-cone frame massive/massles CSF denoted as $\phi_\lcfrm=\phi_\lcfrm (\beta,p^I,u^I)$ depend on the momenta $\beta$, $p^I$ and the unit vector $u^I$, $u^Iu^I=1$. By definition, the massive/massless light-cone frame CSF is constraint-free. The relativistic CSF $\phi$ is expressed in terms of the light-cone frame CSF $\phi_\lcfrm$ as
\be \label{22092025-man03-01}
\phi  = \beta E \phi_\lcfrm\,, \qquad E = \left\{
\begin{array}{ll}
\big(\frac{\xi^+}{\beta}\big)^\SSc &   \hbox{for massive CSF};
\\[9pt]
e^{ \frac{\irm \xi^+}{\beta} }\,, &   \hbox{for massless CSF};
\end{array}\right.
\ee
\!where the variables $p^-$, $\xi^-$, $\xi^I$ entering the relativistic CSF $\phi(p^\mu,\xi^\mu)$ are expressed in terms of  the momenta $\beta$, $p^I$ and the unit vector of the light-cone gauge vector superspace $u^I$, $u^Iu^I=1$,  as
{\small
\beq
\label{22092025-man03-05} && \hspace{-1cm}  p^- = - \frac{p^Ip^I+m^2}{2\beta}\,, \hspace{0.4cm}  \xi^- = -\frac{\xi^I \xi^I}{2\xi^+}\,,    \hspace{1.2cm} \xi^I= \frac{\xi^+}{\beta} \big( |m|u^I + p^I \big),  \hspace{0.5cm} \hbox{for massive CSF};
\nonumber\\[9pt]
&& \hspace{-1cm}  p^- = - \frac{p^Ip^I}{2\beta}\,,  \hspace{1.5cm} \xi^- =  \frac{\kappa^2 - \xi^I\xi^I}{2\xi^+}\,,  \hspace{0.5cm}\xi^I := \kappa u^I + \frac{p^I}{\beta} \xi^+\,,  \hspace{1cm}  \hbox{for massless CSF};
\eeq
}

\noinbf{Relativistic and light-cone gauge massive/massless (triplet) ISF}. Let us use the notation $\phi_\lcrm = \phi_\lcrm (\beta,p^I, \alpha^I,\zeta)$ and  $\phi_\lcrm = \phi_\lcrm (\beta,p^I,\alpha^I)$ for the respective  light-cone gauge massive and massless (triplet) ISF. Then the relativistic ISF $\phi$ given in \eqref{16092025-01-man03-15} is expressed in terms of the light-cone gauge ISF as
{\small
\be \label{22092025-man03-50}
\phi = E \beta \phi_\lcrm \,, \qquad  E := \left\{
\begin{array}{lll}
e^{\alpha^+( \frac{m}{\beta} \zetab - \frac{p^I}{\beta} \alphab^I)}, &    \hbox{for massive (triplet) ISF};
\\[9pt]
e^{- \frac{p^I}{\beta} \alpha^+\alphab^I },   &  \hbox{for massless (triplet) ISF}.
\end{array} \right.
\ee
}
A brutal derivation of \eqref{22092025-man03-50} is as follows. First, one needs to impose the light-cone gauge
{\small
\be \label{22092025-man03-55}
\alphab^+ \phi(p^\mu,\alpha^\mu,\zeta) = 0, \ \ \hbox{for massive (triplet) ISF}; \qquad \alphab^+ \phi(p^\mu,\alpha^\mu) = 0, \ \ \hbox{for massless (triplet) ISF}.
\ee
}
Second, solving the divergence-free constraint \eqref{16092025-man03-16}, we find relations in \eqref{22092025-man03-50}.

As side remark, for the accurate derivation of \eqref{22092025-man03-50}, we should impose a light-cone gauge by taking into account the constraints on the gauge transformation parameters. The accurate derivation also leads us to restrictions \eqref{22092025-man03-50}. In the interest of the brevity, let us skip the details of the accurate derivation. For fields in AdS and conformal fields in flat space, generalization of relation \eqref{22092025-man03-50} may be found in section 4 in ref.\cite{Metsaev:2013kaa}.

\noinbf{Integration measure in light-cone frame}. For the use in appendix \ref{proof}, we consider manifestly Lorentz invariant integration measure for one massive/massless CSF given by
\beq
\label{22092025-man03-10} && d\Gamma  : = \left\{
\begin{array}{ll}
2|m|^{4-d} \delta(p^2+m^2)\delta(p\xi)\delta(\xi^2)d^d p d^d \xi\,,  &  \hbox{for massive CSF};
\\[9pt]
2\kappa^{4-d}\delta(p^2)\delta(p\xi)\delta(\xi^2-\kappa^2)d^d p d^d \xi\,, & \hbox{for massless CSF};
\end{array}\right.
\eeq
Using relations in \eqref{22092025-man03-05}, the measure \eqref{22092025-man03-10} can be cast into the form
\be \label{22092025-man03-15}
d\Gamma =    d\Gamma_{\xi^+}\, d\Gamma_{\xi^-}\,d\Gamma_{p^-}\, d\Gamma_\lcfrm  \,,\qquad  d\Gamma_\lcfrm := \delta(1 -u^2) d\beta d^{d-2}p d^{d-2}u\,,
\ee
where
\beq
\label{22092025-man03-20} && d\Gamma_{p^-}:= \delta\big(p^- + \frac{p^Ip^I+m^2}{2\beta}\big) dp^-\,, \qquad d\Gamma_{\xi^-}:= \delta\big(\xi^-+\frac{\xi^I\xi^I}{2\xi^+}\big) d\xi^-\,,
\nonumber\\
&& d\Gamma_{\xi^+}:= \frac{1}{\xi^+\xi^+} \Big|\frac{\xi^+}{\beta}\Big|^{d-2} d\xi^+\,, \hspace{6cm} \hbox{for massive CSF};
\nonumber\\
&& d\Gamma_{p^-}:= \delta\big(p^- + \frac{p^Ip^I}{2\beta}\big) dp^-\,, \qquad d\Gamma_{\xi^-}:= \delta\big(\xi^-+\frac{\xi^I\xi^I-\kappa^2}{2\xi^+}\big) d\xi^-\,,
\nonumber\\
&& d\Gamma_{\xi^+}:= \frac{1}{\beta^2} d\xi^+\,, \hspace{8cm} \hbox{for massless CSF}.\qquad
\eeq
Note that the light-cone gauge integration measure $d\Gamma_\lcfrm$ \eqref{22092025-man03-15} takes the same form for massive and massless CSFs. For massive CSF, relation \eqref{17092025-man03-30} is given for the domain $\beta>0$, $\xi^+ > 0$.

\section{Proof of relation \eqref{17092025-man03-36}}\label{proof}

In light-cone frame, we obtain the following representation for the various ingredients entering action \eqref{17092025-man03-01}:
\beq
\label{26092025-man03-01} && \Phi_\smp3 =    \Phi_{\smp3,\lcfrm} \prod_{a_\csfsm} \beta_{a_\csfsm} E_{a_\csfsm}\,,
\\
\label{26092025-man03-02} && \LL_\smp3 =   \LL_\lcfrm \prod_{a_\csfsm} E_{a_\csfsm}\,,
\\
\label{26092025-man03-03} && d\Gamma_\smp3 = d\Gamma_{\smp3,\lcfrm} \prod_{a_\csfsm} d\Gamma_{\xi_{a_\csfsm}^+} d\Gamma_{\xi_{a_\csfsm}^-} d\Gamma_{p_{a_\csfsm}^-} \,,
\eeq
where the factor $E$ appearing in \eqref{26092025-man03-01}, \eqref{26092025-man03-02} is defined in \eqref{22092025-man03-01}, while the measures $d\Gamma_{\xi^+}$, $d\Gamma_{\xi^-}$, $d\Gamma_{p^-}$ appearing in \eqref{26092025-man03-03} are given in \eqref{22092025-man03-20}. The product of the three light-cone frame fields $\Phi_{\smp3,\lcfrm}$ in \eqref{26092025-man03-01}, the light-cone frame cubic vertex $\LL_\lcfrm $ in \eqref{26092025-man03-02}, and the light-cone frame integration measure $d\Gamma_{\smp3,\lcfrm}$ in \eqref{26092025-man03-03} are given by the relations
\beq
\label{26092025-man03-08} && \Phi_{\smp3,\lcfrm}= \prod_{a_\csfsm} \phi_{\lcfrm,a_\csfsm}  \prod_{a_\isfsm} \phi_{a_\isfsm}\,,
\\
\label{26092025-man03-09} && \LL_\lcfrm = \LL_\lcfrm (\beta_{a_\csfsm},p_{a_\csfsm}^I,u_{a_\isfsm}^I, p_{a_\isfsm}^\mu,\alpha_{a_\isfsm}^\mu, \zeta_{a_\isfsm})\,,
\\
\label{26092025-man03-10x} && d\Gamma_{\smp3,\lcfrm} = (2\pi)^d \delta^d(\sum_{a=1,2,3} p_a) \prod_{a_\csfsm} d\Gamma_{\lcfrm, {a_\csfsm}} \prod_{a_\isfsm} d\Gamma_{a_\isfsm}\,,
\eeq
where the light-cone frame CSF $\phi_{\lcfrm,a_\csfsm} $ \eqref{26092025-man03-08} has been introduced in \eqref{22092025-man03-01}. In the \eqref{26092025-man03-01}-\eqref{26092025-man03-10x}, the sub-indices $a_\csfsm$ and $a_\isfsm$ label the respective CSFs and ISFs entering the cubic vertex.

Let us now make comment on the derivation of relations \eqref{26092025-man03-01}-\eqref{26092025-man03-10x}. For the derivation of \eqref{26092025-man03-01} we used relation \eqref{22092025-man03-01}. Equations for $\LL_\smp3$ \eqref{17092025-man03-16} take the same form as the constraints for CSF given in \eqref{16092025-01-man03-02}. Therefore, solution for $\LL_\smp3$ in \eqref{26092025-man03-02}, \eqref{26092025-man03-09} is obtained by using the same procedure as the one for CSF in appendices B, C in ref.\cite{Metsaev:2025qkr}. Expression for $d\Gamma_\smp3$ \eqref{26092025-man03-03} is obtained by using the definition in \eqref{17092025-man03-05} and relations \eqref{22092025-man03-15}, \eqref{22092025-man03-20}.

Using \eqref{26092025-man03-01}-\eqref{26092025-man03-10x}, we find the following expression for the integrand in \eqref{17092025-man03-01}:
\be
d\Gamma_\smp3 \Phi_\smp3^\dagger\,\LL_\smp3 =  d\Gamma_{\smp3,\lcfrm} \Phi_{\smp3,\lcfrm}^\dagger\,\LL_\lcfrm  \prod_{a_\csfsm} d\Gamma_{\xi_{a_\csfsm}^+} d\Gamma_{\xi_{a_\csfsm}^-} d\Gamma_{p_{a_\csfsm}^-} \,,
\ee
which can be represented as
\be \label{26092025-man03-10}
d\Gamma_\smp3 \Phi_\smp3^\dagger\,\LL_\smp3 =  dS_\lcfrm d\sigma_\CSFsm  \prod_{a_\csfsm} d\Gamma_{\xi_{a_\csfsm}^-} d\Gamma_{p_{a_\csfsm}^-}\,, \qquad
dS_\lcfrm:= d\Gamma_{\smp3,\lcfrm} \Phi_{\smp3,\lcfrm}^\dagger\,\LL_\lcfrm \,,
\ee
where $d\sigma_\CSFsm$ is defined in \eqref{17092025-man03-30}. Plugging \eqref{26092025-man03-10} in \eqref{17092025-man03-01} and performing integration over $\xi_{a_\csfsm}^-$, $p_{a_\csfsm}^-$, we get \eqref{17092025-man03-36}.

\section{Proof of Lorentz invariance of $S_\smp3^\chi$ \eqref{17092025-man03-20}}\label{lor-inv}

With exception of $\delta_\CSFsm(\chi)$ \eqref{17092025-man03-21}, the integrand entering $S_\smp3^\chi$ \eqref{17092025-man03-20} is manifestly Lorentz invariant. This implies that in order to prove the Lorentz invariance of $S_\smp3^\chi$ we should show that the response of $S_\smp3^\chi$ under Lorentz transformation of $\delta_\CSFsm(\chi)$ given by
{\small
\beq
\label{21092025-man03-01} && \hspace{-1cm} \delta_{J^{\mu\nu}} S_\smp3^\chi = \int d\Gamma_{\smp3,\covrm}\,  \Big(\delta_{J^{\mu\nu} } \delta_\CSFsm(\chi)\Big)\, \Phi_{\smp3,\covrm}^\dagger \LL_\smp3  \,,
\\
\label{21092025-man03-02} &&  \delta_{J^{\mu\nu} } \delta_\CSFsm(\chi)  = \Jbf^{\mu\nu}\delta_\CSFsm(\chi)\,,
\nonumber\\
&& \Jbf^{\mu\nu} := \sum_{a=1,2,3} J_a^{\mu\nu} \,, \quad J_a^{\mu\nu} = p_a^\mu\partial_{p_a^\nu} - p_a^\nu \partial_{p_a^\mu} + \xi_a^\mu \partial_{\xi_a^\nu} - \xi_a^\nu \partial_{\xi_a^\mu} \,,
\eeq
}
\!is equal to zero. Taking into account explicit expression for $\delta_\CSFsm(\chi)$ \eqref{17092025-man03-21}, it is easy to see that $\delta_{J^{+-}} \delta_\CSFsm(\chi)  = 0$, $\delta_{J^{+I}} \delta_\CSFsm(\chi)  = 0$, $\delta_{J^{IJ}} \delta_\CSFsm(\chi)  = 0$. These relations imply immediately the $J^{+-}$- $J^{+I}$-, and $J^{IJ}$- Lorentz invariance of the action $S_\smp3^\chi$. We now consider $J^{-I}$-Lorentz transformations. To this end using the definition of $ \delta_{J^{-I}} \delta_\CSFsm(\chi)$ given in \eqref{21092025-man03-02} and relations \eqref{22092025-man03-05}, we find
\beq
\label{21092025-man03-05} && \delta_{J^{-I}} \delta_\CSFsm(\chi) =  \sum_a \frac{\xi_a^+}{\beta_a} |m_a| u_a^I \partial_{\xi_a^+} \delta_\CSFsm(\chi),  \hspace{1cm} \hbox{for massive CSF};
\\
\label{21092025-man03-06} &&  \delta_{J^{-I}} \delta_\CSFsm(\chi) = \sum_a  \kappa_a u_a^I \partial_{\xi_a^+} \, \delta_\CSFsm(\chi)\,, \hspace{1.7cm} \hbox{for massless CSF}.
\eeq
Now plugging \eqref{21092025-man03-05}, \eqref{21092025-man03-06} in \eqref{21092025-man03-01}, using the relation \eqref{26092025-man03-10}, and performing integration over $\xi_{a_\csfsm}^-$, $p_{a_\csfsm}^-$, we represent the variation \eqref{21092025-man03-01} as
\be \label{21092025-man03-15}
\hspace{-1cm} \delta_{J^{-I}} S_\smp3^\chi = \int  \Big(\delta_{J^{-I}} \delta_\CSFsm(\chi) \Big) d\sigma_\CSFsm^{\vphantom{5pt}} dS_\lcfrm\,.
\ee
We recall that, for the both massive and massless CSFs, the  $dS_\lcfrm$ is independent of $\xi_a^+$. For massless CSF, the measure $d\sigma_\CSFsm$ is also independent of $\xi_a^+$. Taking into account \eqref{21092025-man03-06}, we see that the integrand in \eqref{21092025-man03-15} is a total derivative of $\xi_a^+$ and hence $ \delta_{J^{-I}} S_\smp3^\chi=0$.
For massive CSF, the measure $d\sigma_\CSFsm$ is proportional to $(\xi_a)^{-1}$, while the $\delta_{J^{-I}} \delta_\CSFsm(\chi)$ \eqref{21092025-man03-05} is proportional to $(\xi_a^+)^1$. This implies that the integrand in \eqref{21092025-man03-15} is again a total derivative of $\xi_a^+$ and hence $ \delta_{J^{-I}} S_\smp3^\chi=0$.

\section{Connection between covariant and light-cone gauge cubic vertices}\label{connect}

The Lorentz covariant vertex $\LL_\smp3$ depends on the variables $p_a^\mu$, $\xi_a^\mu$, $\alpha_a^\mu$, $\zeta_a$.
Let us introduce the following notation for the light-cone gauge projection of these variables:
{\small
\beq
\label{29092025-man03-01} && \hspace{-1cm} \xi^+|_\lcgrm =\xi^+,\quad  \xi^-|_\lcgrm = -\frac{\xi^I \xi^I}{2\xi^+}\,,    \hspace{1.2cm} \xi^I|_\lcgrm= \frac{\xi^+}{\beta} \big( |m|u^I + p^I \big),  \hspace{0.5cm} \hbox{for massive CSF};
\nonumber\\[9pt]
&& \hspace{-1cm} \xi^+|_\lcgrm =\xi^+,\quad \xi^-|_\lcgrm =  \frac{\kappa^2 - \xi^I\xi^I}{2\xi^+}\,,  \hspace{0.5cm}\xi^I|_\lcgrm := \kappa u^I + \frac{p^I}{\beta} \xi^+\,,  \hspace{1.2cm}  \hbox{for massless CSF};
\nonumber\\
&& \hspace{-1cm} \alpha^+|_\lcgrm = 0 \,, \quad \alpha^-|_\lcgrm = - \frac{p^I}{\beta}\alpha^I  + \frac{m}{\beta}\zeta\,, \quad \alpha^I,\zeta|_\lcgrm = \alpha^I,\zeta\,, \hspace{1.6cm} \hbox{for massive (triplet) ISF}
\nonumber\\
&& \hspace{-1cm} \alpha^+|_\lcgrm = 0 \,, \quad \alpha^-|_\lcgrm = - \frac{p^I}{\beta}\alpha^I\,, \hspace{1.5cm} \alpha^I|_\lcgrm = \alpha^I, \hspace{2.4cm} \hbox{for massless (triplet) ISF}
\nonumber\\
&& p^-|_\lcgrm = - \frac{p^Ip^I+m^2}{2\beta}\,, \ \  \hbox{for massive fields} \qquad  p^-|_\lcgrm = - \frac{p^Ip^I}{2\beta}\,,  \ \ \hbox{for massless fields}.
\eeq
}
\! Also, for the references, we introduce the notation
\be \label{29092025-man03-08}
\ccc_a := \frac{|m_a|\xi_a^+}{\beta_a}\,, \quad \hbox{for massive CSF}\,, \qquad
\ccc_a := \kappa_a \,, \quad \hbox{for massless CSF}\,.
\ee
Using notation \eqref{29092025-man03-01}, we then note that the Lorentz covariant vertex $\LL_\smp3$ is connected to the light-cone gauge vertex $\LL_\lcgrm$ by the following relation
{\small
\be  \label{29092025-man03-15}
\LL_\smp3(p_a^\mu,\xi_a^\mu,\alpha_a^\mu,\zeta_a)|_\lcgrm  = \LL_\lcgrm \prod_{a_\csfsm} E_{a_\csfsm}, \qquad \LL_\lcgrm = \LL_\lcgrm(\beta_a\,, p_a^I\,, u_a^I,\alpha_a^I,\zeta_a)\,,
\ee
}
\!where the factor $E$ is defined in \eqref{22092025-man03-01}, while $\LL_\lcgrm$ is the light-cone gauge cubic vertex studied in ref.\cite{Metsaev:2025qkr}. In  ref.\cite{Metsaev:2025qkr} the light-cone gauge cubic vertex was denoted as $p_\smp3^-$. Up to some factor depending on $|m_a|$ and $\kappa_a$,  the $\LL_\lcgrm$ is equal to $p_\smp3^-$. Using explicit relations for the $\LL_\smp3$ obtained in this paper and explicit expressions for the light-cone gauge vertex obtained in ref.\cite{Metsaev:2025qkr} we verified that the relation holds \eqref{29092025-man03-15} true. See our comments in sections \eqref{sec-3csf}-\eqref{sec-1csf}. Needless to say that, in \eqref{29092025-man03-15}, we should keep in mind the energy-momentum conservation law governed by the delta-function $\delta^d(\sum_{a=1,2,3}p_a^\mu)$.


\end{document}